\documentclass[sn-nature]{sn-jnl}

\usepackage{graphicx}%
\usepackage{multirow}%
\usepackage{amsmath,amssymb,amsfonts}%
\usepackage{amsthm}%
\usepackage{mathrsfs}%
\usepackage[title]{appendix}%
\usepackage{xcolor}%
\usepackage{textcomp}%
\usepackage{manyfoot}%
\usepackage{booktabs}%
\usepackage{algorithm}%
\usepackage{algorithmicx}%
\usepackage{algpseudocode}%
\usepackage{listings}%
\usepackage{soul}%
\usepackage{hyperref} 
\usepackage{CJKutf8}

\theoremstyle{thmstyleone}%
\theoremstyle{thmstyletwo}%

\theoremstyle{thmstylethree}%

\begin{document}

\title[Article Title]{A $\wedge$-like metallicity profile in the disk of the Sagittarius dwarf galaxy}


\author [1,2]{\fnm{Zixi} \sur{Guo~\text{\begin{CJK*}{UTF8}{gbsn}(郭子熙)\end{CJK*}}}}
\equalcont{These authors contributed equally to this work.}

\author*[1,3]{\fnm{Xiang-Xiang} \sur{Xue~\text{\begin{CJK*}{UTF8}{gbsn}(薛香香)\end{CJK*}}}}\email{xuexx@nao.cas.cn}
\equalcont{These authors contributed equally to this work.}

\author[4]{\fnm{Jiang} \sur{Chang~\text{\begin{CJK*}{UTF8}{gbsn}(常江)\end{CJK*}}}}

\author*[1,2,3]{\fnm{Chao} \sur{Liu~\text{\begin{CJK*}{UTF8}{gbsn}(刘超)\end{CJK*}}}}\email{liuchao@nao.cas.cn}

\author[2]{\fnm{Zehao} \sur{Zhong~\text{\begin{CJK*}{UTF8}{gbsn}(钟泽昊)\end{CJK*}}}}

\author[1]{\fnm{Bingqing} \sur{Zhang~\text{\begin{CJK*}{UTF8}{gbsn}(张冰清)\end{CJK*}}}}

\author[1,5,6]{\fnm{Ruizhi} \sur{Zhang~\text{\begin{CJK*}{UTF8}{gbsn}(张睿之)\end{CJK*}}}}

\author[1,7]{\fnm{Shi} \sur{Shao~\text{\begin{CJK*}{UTF8}{gbsn}(邵实)\end{CJK*}}}}

\author*[1,2,8]{\fnm{Gang} \sur{Zhao~\text{\begin{CJK*}{UTF8}{gbsn}(赵刚)\end{CJK*}}}}\email{gzhao@nao.cas.cn}

\author[8]{\fnm{Zhuohan} \sur{Li~\text{\begin{CJK*}{UTF8}{gbsn}(李卓翰)\end{CJK*}}}}

\author[9,4]{\fnm{Xi} \sur{Kang~\text{\begin{CJK*}{UTF8}{gbsn}(康熙)\end{CJK*}}}}%

\author[1,2,3]{\fnm{Jie} \sur{Wang~\text{\begin{CJK*}{UTF8}{gbsn}(王杰)\end{CJK*}}}}%

\author[10]{\fnm{Jianhui} \sur{Lian~\text{\begin{CJK*}{UTF8}{gbsn}(连建辉)\end{CJK*}}}}

\author[1]{\fnm{Haining} \sur{Li~\text{\begin{CJK*}{UTF8}{gbsn}(李海宁)\end{CJK*}}}}

\author[1,2]{\fnm{Ruizheng} \sur{Jiang~\text{\begin{CJK*}{UTF8}{gbsn}(江睿铮)\end{CJK*}}}}

\affil[1]{\orgdiv{National Astronomical Observatories}, \orgname{Chinese Academy of Sciences}, \orgaddress{ \city{Beijing}, \postcode{100101}, \country{China}}}

\affil[2]{\orgdiv{School of Astronomy and Space Science}, \orgname{University of Chinese Academy of Sciences}, \orgaddress{ \city{Beijing}, \postcode{100049}, \country{China}}}

\affil[3]{\orgdiv{Institute for Frontiers in Astronomy and Astrophysics}, \orgname{Beijing Normal University}, \orgaddress{ \city{Beijing}, \postcode{102206}, \country{China}}}

\affil[4]{\orgdiv{Purple Mountain Observatory}, \orgname{Chinese Academy of Sciences}, \orgaddress{ \city{Nanjing}, \postcode{210023}, \country{China}}}

\affil[5]{\orgname{Instituto de Astrofísica de Canarias}, \orgaddress{ \city{La Laguna}, \postcode{38205}, \country{Spain}}}

\affil[6]{\orgdiv{Departamento de Astrofísica}, \orgname{Universidad de La Laguna}, \orgaddress{\city{La Laguna}, \postcode{38205}, \country{Spain}}}

\affil[7]{\orgdiv{Institute of Astronomy and Physics}, \orgname{Inner Mongolia University}, \orgaddress{\city{Hohhot}, \postcode{010021}, \country{China}}}

\affil[8]{\orgdiv{Shandong Key Laboratory of Space Environment and Exploration Technology}, \orgdiv{Institute of Space Sciences}, \orgname{Shandong University}, \orgaddress{\city{Weihai}, \postcode{264209}, \country{China}}}

\affil[9]{\orgdiv{Institute for Astronomy, the School of Physics}, \orgname{Zhejiang University}, \orgaddress{ \city{Hangzhou}, \postcode{310027}, \country{China}}}

\affil[10]{\orgdiv{South-Western Institute for Astronomy Research}, \orgname{Yunnan University}, \orgaddress{ \city{Kunming}, \postcode{650091}, \country{China}}}


\abstract{
 Unveiling the properties of the progenitors that formed the tidal structures of the Milky Way (MW) is essential for exploring the MW's assembly history. Such studies further afford a unique opportunity to dissect the formation and evolution of dwarf galaxies by contrasting surviving and disrupted systems. However, reconstructing the intrinsic structure of progenitors solely from the positions and velocities of their tidal debris remains a formidable challenge. Here, we decode the chemo-kinematic signatures embedded in the stellar streams of the Sagittarius (Sgr) dwarf galaxy through N-body simulations. Our simulation reproduces a vertical metallicity gradient in the Sgr streams. We show that this feature is naturally explained if the stream originated from a disk component with a $\wedge$-like metallicity profile. Remarkably, similar metallicity profiles are identified in MaNGA survey data for a few (3/217) field dwarf galaxies, suggesting this feature is rare but not unique to Sgr. The vertical metallicity gradient of stellar streams preserves extra detailed galactic fossil records, enabling quantitative reconstruction of the peculiar structure of the satellite galaxy. The similar metallicity profiles of Sgr and MW raise the question of a possible special coevolution. 
}



\maketitle
\section*{Article}
Reconstructing the progenitors of disrupted dwarf galaxies from their stellar tidal streams is fundamental to unraveling the Milky Way's accretion history. This approach not only identifies the building blocks of our Galaxy but also uniquely captures the pre-disruption structure and chemical history of these dwarf galaxies. Recovering this pre-accretion snapshot provides crucial constraints on the formation and early evolutionary processes of dwarf galaxies, complementing observations of their surviving counterparts.
The Sagittarius streams\cite{ibata_dwarf_1994,ibata_archer_1998,majewski_two_2003,belokurov_field_2006,koposov_sagittarius_2012,Slater_2013,deboer_2015,Luque_2017,Navarrete_2017,bennett_exploring_2022}, the most prominent tidal structures in the MW resulting from the ongoing tidal disruption and accretion of the Sgr galaxy, are uniquely advantageous for tracing the progenitor in detail. It has been extensively studied using N-body simulations to reproduce the metallicity gradient along the streams and their 6D phase-space distribution, including core, leading/trailing arms, and bifurcations\cite{law_sagittarius_2010,lokas_inner_2010,penarrubia_was_2010, DL2017, vasiliev_tango_2021,Pino_2021,oria_revisiting_2022,Cunningham_2024}. These studies explore the progenitor’s properties, such as its mass and morphology. However, there is still no clear consensus on the presence of a disk in the progenitor. Large spectroscopic surveys, such as the Large Sky Area Multi-Object Fiber Spectroscopic Telescope (LAMOST) and the Sloan Digital Sky Survey (SDSS), combined with the Gaia mission, offer a new opportunity to investigate the progenitor through the distinct chemo-kinematics of the Sgr streams.



By combining spectroscopic catalogs of LAMOST DR8 K-giants\cite{zhang_catalog_2023}, SEGUE K-giants\cite{Xue_segueKG_catalog_2014}, SDSS DR8 blue-horizontal-branch (BHB) stars\cite{Xue_sdssbhb_catalog_2011}, LAMOST DR5 M-giants\cite{Zhong_lamostMG_catalog_2019}, and proper motions from Gaia DR3\cite{gaia_dr3}, we identify $\sim$3900 members of the Sgr streams in the integrals-of-motion (energy $E$ and angular momentum $\vec{L}$) space. These parameters are derived from radial velocities, distances, celestial coordinates, and proper motions, assuming a spherical Galactic potential\cite{Wang_method_2022, Sun_method_2025}. This Sgr stream member catalog is an updated version of ref.\cite{yang_tracing_2019}. 

\begin{figure}[h]
    \centering
    \includegraphics[width=13cm,height=6cm]{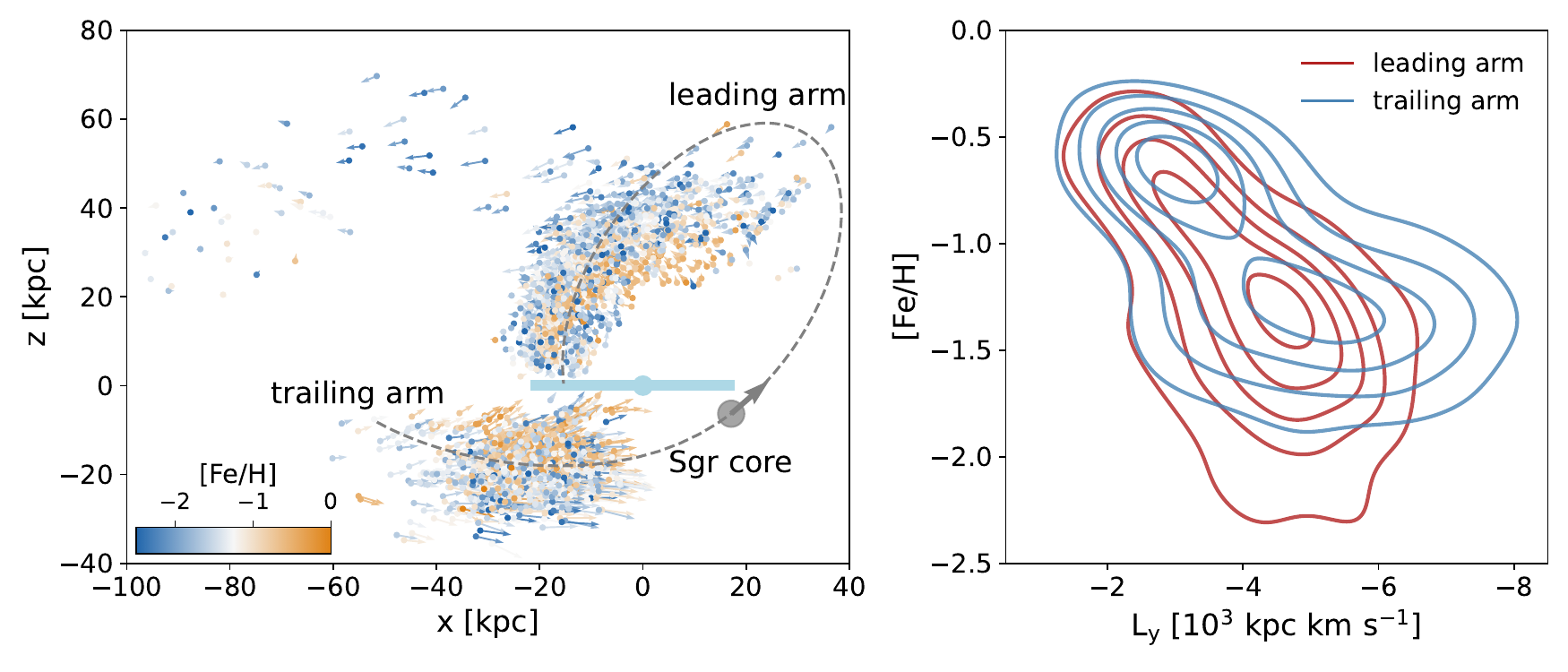}
    \caption{\textbf{Distribution of the Sgr stream members. Left panels: }The spatial distribution of the Sgr leading arm and trailing arm in the Galactocentric coordinate system. Milky Way disk plane is shown by blue line, and its center by a blue dot. The grey dot represents the position of Sgr core, the gray dashed line traces its orbital trajectory, and the grey arrow indicates the direction of its movement. Arrows represent the Sgr stars’ moving directions and velocity amplitudes, and every star is color-coded according to its metallicity. 
    \textbf{Right panel: }${\rm[Fe/H]}-L_y$ correlation. The red contour and the blue contour represent the distribution of the leading arm and the trailing arm, respectively. We can see $L_y$ and ${\rm[Fe/H]}$ have a correlation in both the leading arm and the trailing arm. 
    }
    \label{obs_data}
\end{figure}

The left panel of Fig. \ref{obs_data} presents the spatial distribution of Sgr stream members in the Galactocentric $X-Z$ plane, with velocity vectors indicated by arrows and metallicity (${\rm[Fe/H]}$) encoded in color. Along the direction perpendicular to the motion of the streams, the metal-rich stars of both leading and trailing arms reside closer to the Galactic center than the metal-poor stars. We note that this feature is not caused by the bifurcation (see Supplementary Information Section 3). Given the minimal inclination of the Sgr orbital plane relative to the $X-Z$ plane, $L_y$ naturally constitutes the dominant component of the angular momentum $\vec{L}$. Therefore, to quantitatively describe this vertical metallicity gradient, we utilize $L_y$ as a present-day kinematic metric. We identify a clear ${\rm[Fe/H]}$ gradient along $L_y$ in both arms (the right panel of Fig. \ref{obs_data}). This observed ${\rm[Fe/H]}-L_y$ correlation points to a plausible scenario in which the Sgr progenitor had a rotating disk component.


To investigate whether a disk and what kind of metallicity profile in the Sgr progenitor could reproduce the observed ${\rm[Fe/H]}-L_y$ correlation, we conduct a series of N-body simulations of the Sgr progenitor. These models trace the Sgr progenitor's orbit backward for $\sim$3.6 Gyr, with its orbit rigorously constrained by the present-day positional and kinematic measurements of the remnant. The simulated streams were then forward-evolved and subjected to identical observational selection criteria as the actual data, ensuring a self-consistent comparison between simulations and observations.


Recent observations\cite{2023NatAs...7..951L} reveal that the MW exhibits a $\wedge$-like metallicity profile along the disk. Motivated by these findings, we have constructed six kinds of models to simulate potential configurations of the Sgr progenitor. These models combine three distinct galactic morphologies (disky, spheroidal, and rotating elliptical) with two metallicity profiles: a monotonic decreasing ($\setminus$-like) profile and a $\wedge$-like profile. In the right panel of Fig. \ref{obs_data}, we notice that the trailing arm shows two dense regions in the ${\rm[Fe/H]}-L_y$ map, which might originate from two components in the progenitor. Therefore, in the $\wedge$-like model, we divide the stellar particles into two components. The metallicity of the first component follows a Gaussian distribution with a mean ${\rm[Fe/H]}$ of $-1.3$, and the second component has a metallicity that ascends initially and then descends with the radius.

The mapping between the position of Sgr streams in the Milky Way and their position in the progenitor galaxy is established through N-body simulations of the Milky Way accreting the Sgr dwarf galaxy. As demonstrated in Fig. \ref{obs_data} for the LAMOST and SDSS observed Sgr streams, the leading arm is found to be more distant along the orbit from the Sgr core than the trailing arm. This greater distance along the orbit correlates with earlier stripping times. Consequently, observations show the leading arm predominantly originates from the progenitor’s less-bound outer regions, while the trailing arm stems from the inner region.

\begin{figure*}[h]
    \includegraphics[width=13cm,height=7cm]{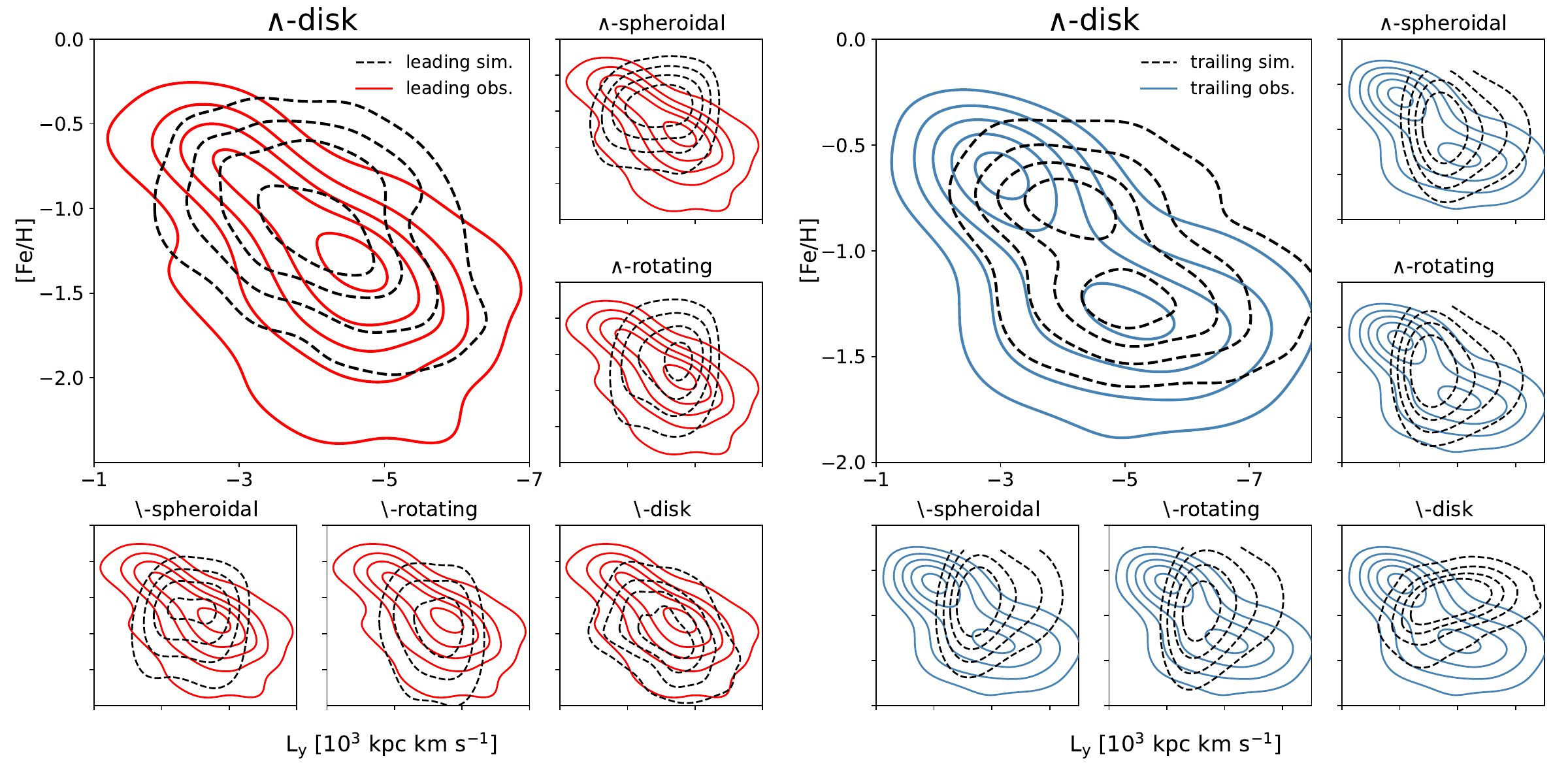}
	\caption{\textbf{ Our simulations revealed the progenitor of Sgr stream. Left panel:} The red and black contours represent the observation data and the mock data of the leading arm. \textbf{Right panel:} The blue and black contours represent the observation data and the mock data of the trailing arm. Among the models, only the $\wedge$-disk model (top-left plot in each panel) can generate the ${\rm[Fe/H]}-L_y$ correlation similar to the observation.
	}\label{test}
\end{figure*}

As shown in Fig. \ref{test}, simulations with both metallicity profiles fail to produce a significant ${\rm[Fe/H]}-L_y$ correlation when spheroidal or rotating elliptical Sgr progenitors undergo tidal disruption. However, in the case of a $\wedge$-like metallicity profile, a disky progenitor can reproduce a ${\rm[Fe/H]}-L_y$ correlation consistent with the observations (shown as Fig. \ref{test}). This finding deviates from the traditional knowledge that the average radial metallicity gradient of a galaxy is generally negative, indicating that metallicity decreases with increasing distance from the galactic center\cite{radial_feh_gradient_Searle_1971,radial_feh_gradient_Zaritsky_1994,radial_feh_gradient_Pilkington_2012,radial_feh_gradient_Sanchez_2014,radial_feh_gradient_manga}.

\begin{figure}[h!]
    \centering
    \includegraphics[width=12cm,height=12cm]{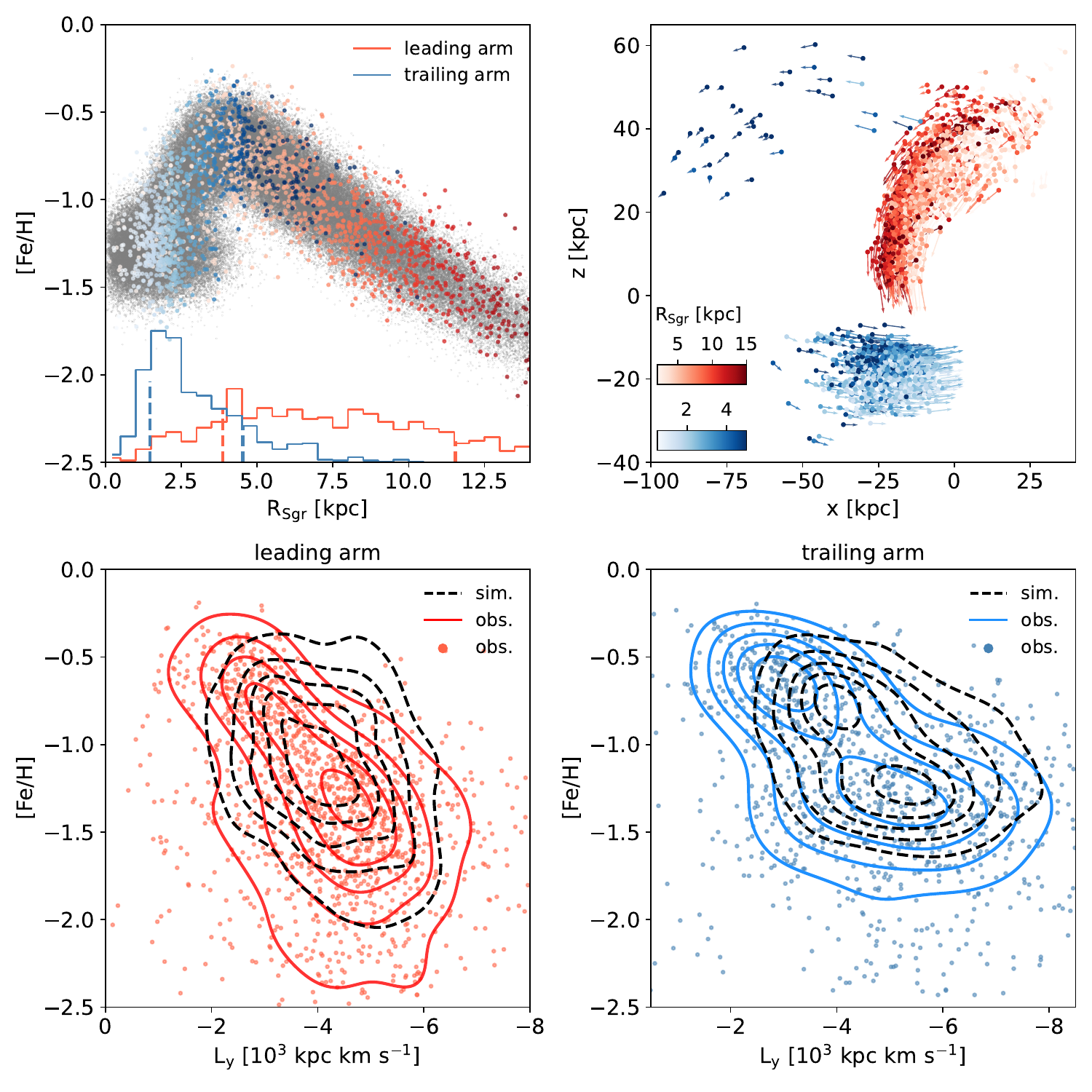}

    \caption{\textbf{Comparison between observation and the fiducial model.} 
    \textbf{Top left:} Initial metallicity profile of the Sgr disk in the simulation. Gray dots represent the distances of stellar particles from the center of Sgr and their metallicities in the initial condition. Red and blue points represent the leading and trailing arms in the mock data, respectively. The two histograms display the initial radial distributions of particles belonging to the leading and trailing arms, with red and blue dashed lines marking the central 68\% interval (16th–84th percentiles) for each distribution. 
    \textbf{Top right:} Spatial distribution of the same leading arm (red) and trailing arm (blue) mock data, with arrows indicating their directions of motion. The parameter $R_{\rm Sgr}$ encodes the distance between particles and the progenitor's center in the initial conditions. Note that the color bar on the right applies to both panels in the top row.
    \textbf{Bottom left:} The red contour and red dots represent the observation data, and the black contour indicates the ${\rm[Fe/H]}-L_y$ correlation of the leading arm in the simulation. 
    \textbf{Bottom right: } The Blue contour and blue dots represent the observation data, and the black contour indicates the ${\rm[Fe/H]}-L_y$ correlation of the trailing arm in the simulation.}
    \label{sim_obs}
\end{figure}

We investigated the vertical structure of both arms by analyzing the initial radii ($R_{\rm Sgr}$) of their constituent particles. The top right panel of Fig. \ref{sim_obs} displays the spatial distribution of mock data from our fiducial model, color-coded by the $R_{\rm Sgr}$. In the direction perpendicular to the stream, the outer side of the leading arm (larger $|L_y|$) is found to contain predominantly particles that originated from larger radii in the progenitor before stripping. In contrast, the opposite happens in the trailing arm. Recall the observations, both the leading and trailing arms contain a higher fraction of metal-poor stars on their outer sides (see Fig. \ref{obs_data}). Therefore, the two arms would show opposite metallicity trends in the progenitor (the top left panel of Fig. \ref{sim_obs}). Consequently, the $\wedge$-like metallicity profile naturally emerges when mapping the chemo-kinematics of the stream back to the Sgr progenitor.

Furthermore, the central $68$\% of the initial radii ($16$th–$84$th percentiles) for particles in the leading arm spans approximately $3.7–11.5$ kpc, whereas the trailing arm particles originate from significantly smaller radii ($1.5–4.5$ kpc). Consequently, our simulations provide meaningful constraints on the metallicity profile of the progenitor's disk across the radial range of approximately $1.5-11.5$ kpc. However, the structure of the innermost region ($\lesssim1.5$ kpc) is less constrained by our data alone. We explore alternative scenarios for the metallicity distribution of the progenitor's core in Supplementary Information Section 4.

As shown in Fig. \ref{sim_obs}, while the mock data for both the leading and trailing arms successfully capture the overall tendency of the observations in the ${\rm[Fe/H]}-L_y$ space, there are slight deviations. These discrepancies primarily stem from our simplifying assumptions, particularly our use of a static Galactic gravitational potential and a limited integration time that traces the Sgr progenitor only through its most recent $\sim$3.5 orbital periods. As a result, the model cannot fully capture the long-term orbital decay driven by dynamical friction. This naturally leads to a narrower distribution of $L_y$ in the mock data compared to the observations. Nevertheless, the overarching trends of the mock data consistently align with the observational data,  supporting the scenario that the Sgr progenitor galaxy hosted a $\wedge$-like disk.

Our exploration of disk models reveals that the observed ${\rm[Fe/H]}-L_y$ correlation is sensitive to several key physical factors: the spin angular momentum orientation, the metallicity profile slopes, the stellar mass, and the disk scale length. The orientation of the progenitor's spin relative to its orbit, defined by the angle $\theta$, plays a critical role in shaping the tidal stream\cite{1972ApJ...178..623T}. We find that the observed ${\rm[Fe/H]}-L_y$ correlation is well reproduced when $\theta$ is in the range of $130^\circ-170^\circ$. This result supports a retrograde rotation scenario, consistent with independent constraints derived from the Sgr bifurcation\cite{penarrubia_was_2010}, although it stands in contrast to recent studies favoring prograde configurations based on the Sgr core or faint branch of the stream\cite{Pino_2021,oria_revisiting_2022}. Beyond the spin orientation, the correlation also shows marked sensitivity to the progenitor's structural parameters. To match the observations, our analysis suggests that the progenitor disk likely had an extended scale length ($R_{\rm s} > 2$ kpc) and a low stellar mass fraction $(M_*/M_{\rm total} < 0.08)$, where $M_*$ is the stellar mass and $M_{\rm total}$ is the total mass of the progenitor galaxy

The initial metallicity profile directly drives the ${\rm[Fe/H]}-L_y$ correlation in the Sgr stream. In turn, reconstructing this $\wedge$-like profile relies on mapping the stream's present-day chemo-kinematics back to the progenitor. Although both leading and trailing arms contain particles from a wide range of initial radii, each arm primarily constrains a different region. The ${\rm[Fe/H]}-L_y$ correlation in the leading arm is dominated by particles stripped from the descending segment of the second component (corresponding to $3.7\ {\rm kpc}\lesssim R_{\rm Sgr}\lesssim 11.5\ {\rm kpc}$). Complementarily, the trailing arm predominantly constrains the ascending segment of the second component (corresponding to $1.5\ {\rm kpc}\lesssim R_{\rm Sgr}\lesssim 4.5\ {\rm kpc}$). Consequently, relatively shallow slopes in these specific segments would significantly weaken the ${\rm[Fe/H]}-L_y$ correlations in the leading and trailing arms, respectively (see Methods Disk model). Furthermore, we verify that the simulated ${\rm[Fe/H]}-L_y$ correlation remains fundamentally robust even when accounting for the external dynamical perturbation introduced by the Large Magellanic Cloud (see Methods Impact of Large Magellanic Cloud and Supplementary Information Section 5).
From the parameter space of disk configurations explored in our simulations, we have identified and present the fiducial model in Fig. \ref{sim_obs}. The inversion of radial metallicity gradient might be attributed to the specific star formation history, but a $\wedge$-like distribution is not extremely rare in the universe. The integrated metallicity profile of the Milky Way is observed to be $\wedge$-like, and there exist Milky Way-mass galaxies with similar radial metallicity distribution in the MaNGA survey and TNG50 numerical simulations\cite{2023NatAs...7..951L}. We have found a few dwarf galaxies with a clearly $\wedge$-like radial metallicity distribution in the MaNGA survey (Fig. \ref{fig04}), which shows that this distribution is also reasonable for dwarf galaxies. The proportion of dwarf galaxies in our sample with a similar distribution is also very low, just over 1\%, which is comparable to the proportion of MW-mass galaxies with such $\wedge$-like metallicity distributions\cite{2023NatAs...7..951L}. Although the specific reason is unknown, this uncommon metallicity profile may indicate that Sgr and the Milky Way, as well as these few galaxies in MaNGA, have experienced similar physical processes, which requires further study and analysis.


\clearpage
\newpage

\section*{Methods}\label{sec2}
\subsection*{Data }\label{subsec2}
\subsubsection*{Observational data}

The identification of the Sgr stream members was performed in the Integrals-of-Motion (I.o.M.) space to overcome the limitations of phase mixing. As shown by ref.\cite{Wang_method_2022,Sun_method_2025,yang_tracing_2019,Penarrubia_2021}, stream members share a global kinematic consensus (i.e. similar orbits), even when their local velocities differ significantly. 

The initial sample used for this clustering consists of K-giant and Blue Horizontal Branch (BHB) stars selected from LAMOST DR8 and SDSS/SEGUE. To minimize disk contamination, we rigorously excluded stars with $|z| < 2$\,kpc, as well as metal-rich stars (${\rm [Fe/H]} > -1.0$) with $2 < |z| < 5$\,kpc, yielding a final catalogue of $\sim$58,000 stars.

Assuming the Galactic halo can be described by a spherical potential, the orbit of each star can be characterized by four I.o.Ms: energy ($E$) and angular momenta ($L_x, L_y, L_z$). Specifically, we adopted a three-component potential consisting of a Hernquist bulge, a sphericalized exponential disk, and an NFW halo. While an axisymmetric potential could theoretically be used, it reduces the strictly conserved I.o.Ms to only $E$ and $L_z$. Maintaining a spherical approximation preserves all four I.o.Ms, which makes substructures significantly more distinguishable. This assumption remains highly robust since the overwhelming majority of the Sgr stream resides in the outer halo.


These conserved quantities can be converted into a 5D orbital space comprising the semi-major axis ($a$), eccentricity ($e$), the direction of the orbital pole ($l_{\rm orbit}, b_{\rm orbit}$), and the direction of the apocenter ($l_{\rm apo}$). Notably, $l_{\rm apo}$ is treated as a quasi-integral because it remains nearly constant over one orbital period. 

The proximity between any two stars in this 5D parameter space is quantified by a normalized orbit-likelihood distance, as detailed in \cite{Sun_method_2025}. Using the Friends-of-Friends (FoF) algorithm, we identified stream members based on their shared orbital properties rather than their instantaneous positions and velocities. The critical linking length for group membership was defined as the value immediately preceding a sharp “jump” in group size.\cite{zhang_2024} Compared with phase space, I.o.M space significantly improves the identification efficiency, especially for streams spanning a wide range of orbital phases.


The heliocentric distances for the three stellar tracers in our sample were derived utilizing established methods appropriate for each population. The K-giants account for $\sim$3600 stars in our total sample, with approximately 77\% from LAMOST. To obtain distances for the K-giants, we applied the robust Bayesian framework introduced by Zhang et al. (2023)\cite{zhang_catalog_2023}, achieving a median distance precision of 16\%. For the M-giants ($\sim$50 stars in our total sample), distance derivations relied on the $(J-K)_0$ color–magnitude relation\cite{Li_2016,Li_2019,Zhong_lamostMG_catalog_2019}, which secures a typical accuracy of 15\%. Finally, since the BHBs ($\sim$250 stars in our total sample) act as standard candles, we inferred their distances utilizing their nearly invariant absolute magnitudes\cite{Xue_sdssbhb_catalog_2011}, reaching a precision better than 10\%. 

Our total sample consists of $\sim$3900 stars in the Sgr stream. To ensure data quality, we retained only stars meeting the following error criteria:
$d_{\rm err}/d<0.2,\ v_{\rm los\_err}<15\ \rm \rm km\ s^{-1},\ [Fe/H]_{err}<0.25\ dex$.
Among the selected samples, $\sim$1300 stars are classified as leading arm and $\sim$1100 stars as trailing arm. Further details regarding the data analysis and sample validation are provided in the Supplementary Information Section 1.



\subsubsection*{Mock data}

Given that observational data contain errors, we cannot make a direct comparison between the observational data and the simulation results. The errors associated with the ${\rm[Fe/H]}-L_y$ correlations primarily stem from the uncertainties in line-of-sight velocity($v_{\rm los}$), distance, and metallicity. In our simulations, the relative distance error of stellar particles is approximately 10\%, which is comparable to that in the observational data\cite{zhang_catalog_2023}. The errors in metallicity and $v_{\rm los}$ are established at $\rm 0.12\ dex$ and $\rm 5\ \rm km\ s^{-1}$, respectively. Subsequently, we will select the stellar particles that closely match the position and velocity of the observation data to serve as mock data from the snapshot. Specifically, we select $\sim$850 simulation particles to represent the trailing arm. Among these, $\sim$330 particles belong to the first component and $\sim$520 to the second component. For the leading arm, we select $\sim$1,150 particles, of which approximately 1,080 belong to the second component (the detailed classification of these components is described in Initial metallicity profile of the Methods). Supplementary Fig. 1 shows the mock data of the fiducial model, and we can discover a vertical metallicity gradient consistent with the observations.


\subsection*{Simulation details}
In this simulation, we use a static Milky Way potential modeled by AGAMA that includes both dark matter and stellar components\cite{agama}. The dark matter halo of the Milky Way is described as a spherically symmetric Navarro-Frenk-White (NFW) potential\cite{Navarra_NFW_1997} with virial mass $\ M_{\rm vir}^{\rm MW}=1\times10^{12}M_{\odot} $ and scale length $r_{\rm s}^{\rm MW}=19\ \rm kpc$. 
The stellar component comprises a Miyamoto-Nagai disk ($6.8\times10^{10}M_{\odot}$) exhibiting scale length $R_{\rm s}^{\rm MW}=3\ \rm kpc$ and scale height $R_{\rm h}^{\rm MW}=0.28\ \rm kpc$\cite{Miyamoto_1975}, coupled with an exponentially truncated power-law bulge ($5\times10^{9}M_{\odot}$) characterized by a power-law exponent of $-1.8$ and a cut-off radius of $1.9\ \rm kpc$\cite{bovy_galpy_2015}.

The current position of Sgr and its line-of-sight velocity in the International Celestial Reference System(ICRS) frame are $({\rm ra,dec},d) =(283.76^{\circ},\ -30.48^{\circ},\ 26.0\ \rm kpc)$ and $v_{\rm los}=142\ \rm km\ s^{-1}$, and its mean proper motion is $\mu_{\alpha}=-2.7\ \rm mas\ yr^{-1}, \mu_{\delta}=-1.35\ mas\ yr^{-1}$\cite{vasiliev_last_2020,li_exploring_2024}. Adopting Solar parameters from\cite{wang_local_2021} ($X=(-8,\ 0,\ 0)\ \rm kpc$, $V=(11.7,\ 220+10.16,\ 7.67)\ \rm km\ s^{-1}$), we derive the Galactocentric position and velocity of the Sgr core as $X_{\rm Sgr}=\ (17.1,\ 2.46,\ -6.33)\ \rm kpc$, $V_{\rm Sgr}=\ (234.53,\ -34.39,\ 205.04)\ \rm km\ s^{-1}$. In our simulation, Sgr will pass through the pericenter four times. We apply the Gauss-Newton iteration method to find the initial coordinate of Sgr $\sim$ 3.6 Gyr ago\cite{vasiliev_tango_2021,sheng_uncovering_2024}.

The disk progenitor is modeled as a composite system containing an exponential stellar disk and a Hernquist-profile dark matter halo\cite{Hernquist_1990}, generated using the GalIC code\cite{yurin_iterative_2014}. Given the ${\rm[Fe/H]}-L_y$ correlation's primary dependence on progenitor rotation, the bulge component is intentionally omitted. The simulations are conducted using Gadget-2 and Gadget-4 with $4\times10^5$ particles\cite{springel_cosmological_2005, Springel_2021}. Detailed parameters of the fiducial model can be found in Supplementary Table 1. 

Following established models for massive Sgr progenitors\cite{lokas_inner_2010,DL2017}, we adopt an initial dark matter halo mass of $2\times10^{10}M_{\odot}$. While such a relatively massive halo could induce early dynamical perturbations on the Milky Way disk\cite{Antoja_2018}, it does not qualitatively affect the stream's chemo-kinematics---the central focus of this study. The halo concentration is set to 10, consistent with the theoretical mass–concentration relation at $z\simeq0.5$\cite{Diemer_2015}. The dark matter halo is represented by $ N_{\rm dm}=200,000$ particles with a softening length of $\epsilon_{\rm dm}=0.1$~kpc.

The initial disk scale length is configured based on \cite{lokas_inner_2010}. As demonstrated in Mass and scale length of Sgr disk in the Methods, more compact disks lead to a weaker ${\rm[Fe/H]}-L_y$ correlation. Furthermore, to ensure the surviving Sgr core mass matches current observations ($\sim10^{8}M_{\odot}$)\cite{vasiliev_last_2020}, we assign an initial disk mass of $8\times10^{8}M_{\odot}$. The stellar disk consists of 200,000 particles in total, with a softening length of $\epsilon_{\rm disk} = 0.02$~kpc. As detailed in Initial metallicity profile of the Methods, the disk particles are further divided into two components based on the adopted metallicity profile: Component 1 contains $N_{\rm disk}\ ({ comp.1}) \simeq 46,000$ particles, while Component 2 contains $N_{\rm disk}\ ({comp.2}) \simeq 154,000$ particles.

Finally, the disk's initial orientation is constrained through a comprehensive ($\theta,\phi$) grid search, where$\phi$ is the azimuthal angle of the disk’s spin angular momentum vector (Spin angular momentum vector in the Methods). By directly confronting the simulated ${\rm[Fe/H]}-L_y$ correlations with observational data, we find that the viable inclination angles are restricted to 
$\theta\in[130^\circ,170^\circ]$. From this favored parameter space, we select $(\theta,\phi)=(150^\circ,40^\circ)$ for our fiducial model.

In order to investigate how the various characteristics of the progenitor galaxy are reflected in the stellar stream, we also simulate with the other models--spheroidal galaxy and rotating elliptical galaxy, which have the same dark matter halo as the fiducial model. The detailed parameters of the models' stellar components are presented in Supplementary Table 2. $a/c$ is the stretch factor, and $a = b$ and $c$ are the axes of some isodensity ellipsoid. A prolate galaxy has $a/c < 1$, while an oblate galaxy has $a/c > 1$. The parameter $k$, 
\[
\langle v_\phi \rangle^2=k^2[\langle v_\phi^2 \rangle-\sigma_R^2]
\]
, is used to delineate $\sigma_\phi^2$. Here, $v_\phi$ is the azimuthal velocity, and $\sigma_R$ and $\sigma_\phi$ are the radial and azimuthal velocity dispersions, respectively, in cylindrical coordinates. If we assume $k$ to be 0, then this galaxy is a non-rotating system.


\subsection*{Disk model}\label{sec3}
In the disk model, we focus on the following key parameters exhibiting dominant control over ${\rm[Fe/H]}-L_y$ correlation:(1) spin angular momentum vector($L_{\rm spin}$), (2) mass and scale length of the disk, (3) initial metallicity profile.

\subsubsection*{Spin angular momentum vector}\label{subsec_spin}
In the Galactocentric rest frame, we generate an initial disk whose $L_{spin,0}$ is in the positive direction of the z-axis, and then we use the following matrices to rotate it.

\begin{equation}
R_x(i)=
\begin{bmatrix}
1 & 0 & 0\\[8pt]
0 & {\rm cos}(i) & {\rm -sin}(i)\\[8pt]
0 & {\rm sin}(i) & {\rm cos}(i)
\end{bmatrix}
\end{equation}
 
\begin{equation}
R_z(i)=
\begin{bmatrix}
{\rm cos}(i) & {\rm -sin}(i) & 0\\[8pt]
{\rm sin}(i) & {\rm cos}(i) & 0\\[8pt]
0 & 0 & 1
\end{bmatrix}
\end{equation}

For example, if we rotate $L_{\rm spin,0}$ around the x-axis by an angle of $i\simeq 100^{\circ}$ to get $L_{\rm spin}$, the orbital angular momentum($L_{\rm orb}$) of Sgr and $L_{\rm spin}$ will be in the same direction. Our study holds that the included angle between $L_{\rm orb}$ and the positive direction of the z-axis is $\alpha=100^{\circ}$. We define the inclination angle subtended by $L_{\rm spin}$ and $L_{\rm orb}$ as $\theta\equiv\arccos\left(
\frac{\boldsymbol L_{\rm orb}\cdot\boldsymbol L_{\rm spin}}
{|\boldsymbol L_{\rm orb}||\boldsymbol L_{\rm spin}|}
\right)$ with a range of $0-180^\circ$ and the azimuthal angle as $\phi$ with a range of $0-360^\circ$\cite{penarrubia_was_2010,oria_revisiting_2022}. $L_{\rm spin}$ can be depicted using the following formula:
\begin{equation}
    L_{\rm spin} = R_{x}(\alpha)R_z(\phi)R_x(\theta)L_{\rm spin,0} 
\end{equation}
For $\theta$ and $\phi$, we conduct 162 simulations with 20-degree increments and determine the optimal parameters of our fiducial model to be $\theta=150^{\circ}$ and $\phi=40^{\circ}$. Supplementary Fig. 2 and Supplementary Fig. 3 show the correlation between ${\rm[Fe/H]}$ and $L_y$ with different $\theta$, and we set $\phi$ at $0^{\circ}$ in these two figures. We establish that $\theta$ exerts significant control over the emergence of the ${\rm[Fe/H]}-L_y$ correlation. This causal relationship originates from $\theta$ which critically governs the tidal interaction strength during galactic mergers. At $\theta \to 0^{\circ}$, enhanced tidal stripping produces extended, filamentary streams through violent disk shocking; Conversely, $\theta \to 180^{\circ}$ configurations yield thicker streams due to moderated interaction strength\cite{1972ApJ...178..623T}. Our simulations show that a thick and metallicity-stratified stellar stream aligns more with the current observational data and $\theta$ can be restricted within the range of $130^{\circ}$ to $170^{\circ}$. Although the disk inclination $\theta$ can be constrained relatively well, we cannot determine the range of the azimuthal angle $\phi$ through the ${\rm[Fe/H]}-L_y$ correlation alone, as their relationship in our simulations shows no clear systematic pattern (See Supplementary Information Section 2 for a detailed discussion on the impact of $\phi$).

\subsubsection*{Initial metallicity profile }\label{subsec_metal}
In this section, we test the influence of initial metallicity profile on the ${\rm[Fe/H]}-L_y$ correlation. Before carrying out the simulation, we take the parameters of the fiducial model as the initial condition and set the metallicity for each particle of the disk within a coordinate system where the $Z$-axis is in the same direction as $L_{\rm spin}$ and the center is at the center of Sgr progenitor. The metallicity of individual particles conforms to the subsequent distribution.  
\begin{equation}\label{imd}
{\rm[Fe/H]} \sim
 \begin{cases}
    N_1(\mu_1,\sigma_1) & comp.1,\ p_1 = {\rm max}\{1-l_z/l_{\rm a},0\}\\
    N_2(\mu_2,\sigma_2) & comp.2,\ p_2 = {\rm min}\{l_z/l_{\rm a},1\}
 \end{cases}
\end{equation}
where,
\begin{equation}
 \mu_2=
 \begin{cases}
     k_1(l_z-l_b)-0.6 & (l_z<l_{\rm b}) \\
     -k_2(l_z-l_b)-0.6 & (l_z\ge l_{\rm b})
 \end{cases}
\end{equation}
Regarding other parameters, we take $\mu_1=-1.3\  {\rm dex},\sigma_1 =0.12\ {\rm dex},\sigma_2=0.1\ {\rm dex} ,l_{\rm a}=1.8\ {\rm [10^2\ kpc\ km\ s^{-1}]} ,\text{and }l_{\rm b}=2\ {\rm [10^2\ kpc\ km\ s^{-1}]}$. $k_1$ and $k_2$ are free parameters. $p_1$ and $p_2$ represent the probabilities that a stellar particle with an angular momentum of $l_z$ in the progenitor coordinate system follows the Gaussian distribution $N_1$ with a probability of $p_1$ and the Gaussian distribution $N_2$ with a probability of $p_2$. From the simulation results, we discover that $k_1$ predominantly influences the trailing arm, whereas $k_2$ mainly affects the leading arm. Supplementary Fig. 4 presents the initial metallicity profiles and the ${\rm[Fe/H]}-L_y$ correlations in the simulations under different $k_1$. It can be observed that the trailing arm has two high-density regions. The metal-poor high-density region is mainly contributed by $comp.1$ of equation \ref{imd}. Another high-density region comes from the stellar particles with high metallicity within a radius of 5 $\rm kpc$. Therefore, a small $k_1$ would result in an indistinct ${\rm[Fe/H]}-L_y$ correlation. In Supplementary Fig. 5, we only vary the value of $k_2$. We find that if $k_2$ is too small, the leading arm in the simulation will lack metal-poor stellar particles. Based on the observation and simulation data, we constrain that $k_1$ must be greater than 0.26, and $k_2$ must be greater than 0.15. In our fiducial model, we adopted $k_1=0.36$ and $k_2=0.22$.
 
\subsubsection*{Mass and scale length of Sgr disk }\label{subsub1}
To impose better constraints on the disk of Sgr, we examine the influence of scale length and mass on the ${\rm[Fe/H]}-L_y$ correlation. When we test the impact of the scale length, the disk models are initialized with scale lengths of 1.5 $\rm kpc$, 2 $\rm kpc$, 2.5 $\rm kpc$, and 3 $\rm kpc$. Since the parameters $l_{\rm a}$ and $l_{\rm b}$ dictate the emergence of the ${\rm[Fe/H]}-L_y$ correlation, we apply minor adjustments to them for each specific model. Meanwhile, all other parameters are kept identical to the fiducial model, with $k_1$ and $k_2$ fixed at 0.36 and 0.22, respectively. Supplementary Fig. 6 indicates that the trailing arm will lack metal-poor components due to the smaller disk. Under the same mass, a smaller disk would lead to a deeper gravitational potential in the central region, making it more difficult for the central metal-poor particles to escape. Consequently, there will also be a deficiency of metal-poor stellar particles in its trailing arm. 

In the subsequent test, we exclusively adjust the stellar mass, whereas all other parameters are held constant. From Supplementary Fig. 7, it can be inferred that a greater stellar mass would also give rise to a deficiency of metal-poor components in the trailing arm. According to our simulation results, either a significantly short scale length or an excessively large stellar mass is insufficient to produce a ${\rm[Fe/H]}-L_y$ correlation. Consequently, we contend that in our simulation, the scale length of the Sgr disk should exceed 2$\rm kpc$, and the proportion of stellar mass to total mass should remain below 0.08.



\subsection*{Impact of Large Magellanic Cloud}\label{subsec6}

To validate our results against the time-dependent potential and the MW's reflex motion, we performed additional simulations incorporating the LMC. We adopted the methodology of Vasiliev et al. (2021)\cite{vasiliev_tango_2021} to construct a time-dependent MW+LMC potential. For the density of LMC, following \cite{vasiliev_tango_2021}, we adopted a truncated spherical NFW model with $r_{\rm  t}^{\rm LMC}=10\ r_{\rm h}^{\rm LMC}$.
\begin{equation}
    \rho_{\rm LMC}\propto r^{-1}(1+r/r_{\rm h}^{\rm LMC})^{-2}{\rm exp}[-(r/r_{\rm  t}^{\rm LMC})^2].
    \label{eq_5}
\end{equation}
We adopt a total mass of $M_{ \rm LMC}=1\times10^{11}M_\odot$ and a scale length of $r_{\rm h}^{\rm LMC}=7.8\ \rm kpc$, constrained by requiring the circular velocity at 8.7 kpc to be approximately 92 $\rm km\ s^{-1}$\cite{van_2014}.

We reconstructed the LMC's orbit backward in time by applying the Chandrasekhar prescription for dynamical friction\cite{Chandra_1943}, where the acceleration is given by:
$$ \mathbf{a}_{\rm df} = -4\pi\ G^2\ M_{\rm LMC}\ \rho_{\rm MW}(r)\ \ln\Lambda\ \frac{\mathbf{v}}{v^3} \left[ \mathrm{erf}(X) - \frac{2X}{\sqrt{\pi}} e^{-X^2} \right] $$
where $\ln(\Lambda)$ is the Coulomb logarithm and $X=v/(\sqrt{2}\sigma_{\rm h})$ is the ratio of the LMC's velocity to the velocity dispersion of the MW's dark matter halo. The results for a representative model ($\theta=130^\circ, \phi=0^\circ$) under the combined MW+LMC potential are presented in Supplementary Fig. 8.

\subsection*{Searching dwarf galaxies with $\wedge$-like metallicity profile from MaNGA survey }
The Mapping Nearby Galaxies at Apache Point Observatory (MaNGA) is one of the state-of-art widely used galaxy surveys\cite{2015ApJ...798....7B}, which provides spatially resolved spectra that includes both the kinematics and chemical information of galaxies. The \texttt{pipe3D} pipeline perform a continuum segmentation binning method to produce higher-S/N average spectra\cite{2016RMxAA..52..171S}, which have been used to analyze stellar population for more than 10,000 galaxies in the MaNGA survey\cite{2022ApJS..262...36S}. We made use of the spatially resolved stellar population information from \texttt{pipe3D} value added catalogues to search dwarf galaxies with \(\wedge\)-like metallicity distribution. The dwarf galaxies here refer to galaxies with logarithmic stellar mass \(\log(M/M_{\odot})\) less than 9.5. Based on this, we utilised dwarf galaxies from MaNGA survey with logarithmic stellar mass less than or equal to \(9.5-\log(2)\) within the petrosian r-band effective radius \(R_{e}\), simply assuming that the mass-to-light ratios of these galaxies are constant from the inside to the outside. In addition, to ensure the observed field of view (FoV) can cover most area of galaxies, we only utilised dwarf galaxies with the diagonal radius of the integral field unit (IFU) cube larger than or equal to \(2R_{e}\) \(({\rm FoV }\geq2R_{e})\). Moreover, we only used the observation of 9l and 127 fibers to ensure sufficient data points for the sample. After these steps, we finally got a sample of 217 dwarf galaxies.

\begin{figure*}[h!]
    \includegraphics[width=12cm,height=4cm]{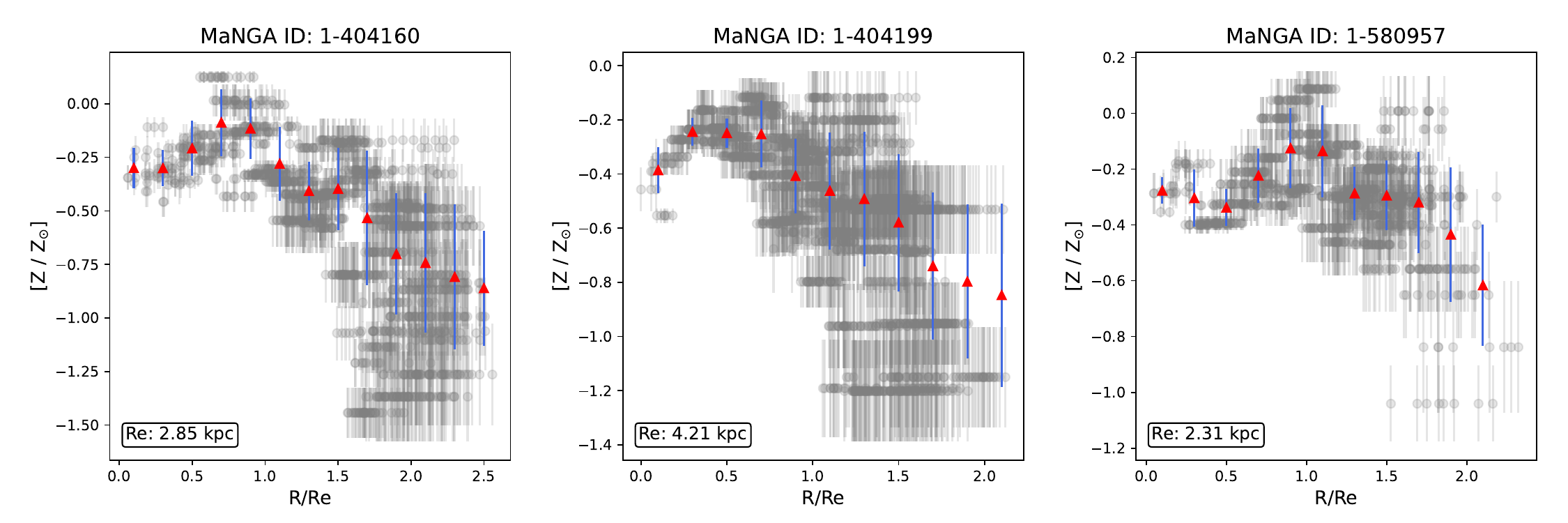}

	\caption{\textbf{Three MaNGA dwarf galaxies with clearly $\wedge$-like radial metallicity distribution.} The radial bin values (red triangles) represent the flux-weighted average of \(\rm{Z}\) in concentric elliptical annuli spaced at 0.2 \(R_{\rm e}\) intervals centered on the galaxy’s optical center. Here \(R_{\rm e}\) is the petrosian r-band effective radius. The error bars of radial bins are contributed by the error propagation of each spaxel and the standard deviation within the elliptical annuli.  
	}\label{fig04}
\end{figure*}

Among this sample, we used the spatially resolved light-weighted metallicity map of each dwarf galaxy to calculate the radial metallicity distribution. As shown in Fig. \ref{fig04}, after binning the scatter data points, only about two or three red triangles are within  \(0.5R_{\rm e}\).We considered a \(\wedge\)-like profile unreliable if its peak occurred within 0.5 $R_{\rm e}$, because such a pattern could arise from excessive randomness. Furthermore, the data points of these 217 samples exhibit larger significant dispersion beyond approximately  \(1.5R_{\rm e}\), also can be seen in this figure. Therefore, we conservatively chose galaxies with \(\wedge\)-like distribution peaks between \(0.5R_{\rm e}\) and \(1.5R_{\rm e}\). In addition, we tend to exclude edge-on dwarf galaxies (Here refers to the galaxies with the axial ratios \(B/A\leq0.5\)), because they have less IFU spaxels at the same stellar mass, and their \(\wedge\)-like metallicity distributions are more difficult to confirm. We excluded some plausible samples, and finally selected three dwarf galaxies with clearly \(\wedge\)-like metallicity distribution by visual inspection, as shown in Fig. \ref{fig04}. Therefore, the overall selection rate from 217 samples was slightly greater than 1\%.

If different selection strategies were employed, or if these plausible samples were taken into account, the corresponding proportion of galaxies with \(\wedge\)-like metallicity distribution could be a few of percentage points, slightly higher than our conservative estimate here. However, we considered this would not make a significant difference, as such galaxies remain very rare.

\section*{Data availability}
M-giants used in this paper are publicly available from \url{https://cdsarc.cds.unistra.fr/viz-bin/cat/J/ApJ/886/154} \cite{yang_tracing_2019}. The catalogue and simulation data are hosted at \url{https://zenodo.org/records/21392279}\cite{guo_2026_15239244}. The data of pipe3D value-added catalogues of MaNGA survey used for this study are publicly available at \url{https://data.sdss.org/datamodel/files/MANGA_PIPE3D/MANGADRP_VER/PIPE3D_VER/SDSS17Pipe3D.html}

\section*{Code availability}
The GADGET code for N-body simulation is available at \url{https://wwwmpa.mpa-garching.mpg.de/galform/gadget/}. The GALIC code for the construction of N-body galaxy models in collisionless equilibrium is available at \url{https://wwwmpa.mpa-garching.mpg.de/\~volker/galic/}. The plot code used in the main text is hosted at \url{https://zenodo.org/records/15239244}\cite{guo_2026_15239244}. 

\section*{Acknowledgements}

Guoshoujing Telescope (the Large Sky Area Multi-Object Fiber Spectroscopic Telescope LAMOST) is a National Major Scientific Project built by the Chinese Academy of Sciences. Funding for the project has been provided by the National Development and Reform Commission. LAMOST is operated and managed by the National Astronomical Observatories, Chinese Academy of Sciences.

We have made use of data from the European Space Agency (ESA) mission
{\it Gaia} (\url{https://www.cosmos.esa.int/gaia}), processed by the {\it Gaia}
Data Processing and Analysis Consortium (DPAC,
\url{https://www.cosmos.esa.int/web/gaia/dpac/consortium}). Funding for the DPAC
has been provided by national institutions, in particular the institutions
participating in the {\it Gaia} Multilateral Agreement.

\section*{Funding}
This study is supported equally by the National Natural Science Foundation of China (12588202) and the National Key R\&D Program of China (2024YFA1611902).
This study is also supported by the National Key R\&D Program of China Nos.2024YFA1611903, 2023YFE0107800,2023YFA1608100, the National Natural Science Foundation of China under grant Nos.12373020, 12573024, 12533007, 12503023, 12090041, 12273053, 12273027, the Strategic Priority Research Program of Chinese Academy of Sciences grant Nos. XDB1160102, the CAS Project for Young Scientists in Basic Research grant No. YSBR-062 and No. YSBR-092, the science research grants from the China Manned Space Project with NO. CMS-CSST-2025-A11, CMS-CSST-2025-A20, and the fellowship from the China Postdoctoral Science Foundation No. 2024M763214.

\section*{Author contributions}

Z.-X. G. contributed most of the modelling and provided simulation data. X.-X. X. proposed and initiated this study. C. L. and X.-X. X. developed the initial idea. G. Z. contributed to the research support. X.-X. X. led the discussions, oversaw the project's progress, and was responsible for interpreting the observational data and theoretical simulations. J. C. taught Z.-X. G. to construct the N-body simulation. X.-X. X. and R.-Z. Z. selected the Sagittarius members from LAMOST DR8. B.-Q. Z. and Z.-H. Z. analyzed the data of the MaNGA survey. As Z.-X. G.’s supervisor, G. Z. provided valuable guidance through regular discussions. S. S. contributed suggestions regarding parameter testing. H.-N. L., J.-H. L., and R.-Z. J. offered suggestions regarding the chemical evolution. X. K. and J. W. advised on the initial conditions of the simulation. Z.-H. L. provided suggestions regarding the initial position of Sgr. Z.-X. G., X.-X. X., Z.-H. Z., and B.-Q. Z. co-wrote the initial manuscript. Z.-X. G., X.-X. X., Z.-H. Z., and R.-Z. Z. co-revised the manuscript. All authors discussed and commented on the manuscript.
\section*{Additional information}

{\bf Correspondence and requests for materials} should be addressed to Xiang-Xiang Xue, Chao Liu, or Gang Zhao.

\section*{Competing interest}
The authors declare no competing interests.

\clearpage
\newpage

\section*{Figure Legends/Captions}

\clearpage
\newpage


\clearpage 
\renewcommand{\figurename}{Supplementary Figure} 
\setcounter{figure}{0}
\renewcommand{\thefigure}{\arabic{figure}}

\renewcommand{\theHfigure}{supplement.\arabic{figure}}
\section*{Supplementary Information}
\section{Observation data}

Supplementary Figure \ref{fig_derr_bin} shows the distribution of relative distance errors in our sample, supporting our adopted quality threshold of $\sigma_d/d < 0.2$. Furthermore, to evaluate the influence of these data cuts on our main findings, Supplementary Figure \ref{fig_derr_fe_ly} compares the ${\rm[Fe/H]}-L_y$ distribution before and after applying the criteria. Specifically, the top row displays the total sample, whereas the bottom row presents the filtered sample restricted by $d_{\rm err}/d < 0.2$, $v_{\rm los, err} < 15\ {\rm km\,s^{-1}}$, and ${\rm [Fe/H]}_{\rm err} < 0.25\ {\rm dex}$. As shown, while the quality cuts effectively reduce observational scatter, the overarching ${\rm[Fe/H]}-L_y$ trend remains consistent. This indicates that the observed metallicity gradient is a robust feature of the dataset and is not driven by the selection process.

To rigorously verify that the observed metallicity gradient is not an artifact of observational selection biases, we performed robustness checks. 

To mitigate potential completeness biases related to absolute magnitude and metallicity, we tested a strictly defined color-magnitude window. The left panel of Supplementary Figure \ref{fig_mag_cut} presents the Color-Magnitude Diagram (CMD) for our sample, from which we isolated a uniform subset of stars confined to $-1.5 < M_G < 1.0$ and $0.35 < (G-G_{\rm RP}) < 0.85$. When restricting our analysis to this narrow range, the ${\rm[Fe/H]}-L_y$ correlation remains clearly visible in both the leading and trailing arms (middle and right panels of Supplementary Figure \ref{fig_mag_cut}). 

To simultaneously assess the impact of tracer-specific biases and distance-dependent selection effects, we performed a spatially resolved analysis of the sample. In Supplementary Figure \ref{tracer_distance}, we partitioned the data for the leading and trailing arms into heliocentric distance ($d$) bins with widths of 10 kpc and 5 kpc, respectively, while explicitly distinguishing between the different tracer populations (K-giants, M-giants, and BHBs).

Although M-giants and BHBs preferentially populate the metal-rich and metal-poor extremes respectively, the sample is overwhelmingly dominated by K-giants, which span the full metallicity range ($\rm -2.5\lesssim [Fe/H]\lesssim 0$). We find that the ${\rm[Fe/H]}-L_y$ correlation is robustly recovered within the K-giant subsample alone, ruling out artifacts from mixing populations with distinct selection functions. Furthermore, within the narrow distance slices where selection probabilities are uniform, the correlation persists across all distance bins for the leading arm and remains significant in the 20–30 kpc interval for the trailing arm. This decoupling confirms that the reported chemo-kinematic gradient is an intrinsic structural feature of the stream and not a consequence of distance-dependent selection.

Furthermore, to demonstrate that this correlation is maintained along the longitudinal extent of the stream, we examined the observed sample within distinct $\tilde{\Lambda}_\odot$ bins (see Supplementary Figure \ref{fig_lambda_bin}). We confirm that the clear ${\rm[Fe/H]}-L_y$ correlation is robustly present locally within the well-sampled regions of both the leading and trailing arms. This longitudinal persistence, combined with the distance-resolved analysis, conclusively demonstrates that the vertical metallicity gradient is an intrinsic, local structural feature of the Sagittarius stream.

\section{Simulation}
\subsection{Orbit and Initial Mass}
In our fiducial model, we adopt a relatively high initial total mass  ($2\times 10^{10}\,M_\odot$) for the Sgr progenitor. Massive progenitors of this scale have been widely utilized in previous literature to explore the Milky Way-Sgr interaction (e.g., an initial mass of $\sim 1.6\times 10^{10}\,M_\odot$ in \cite{lokas_inner_2010}, and $\sim 10^{10}\,M_\odot$ in \cite{DL2017}). 

The primary difference between our model and some recent low-mass setups (Vasiliev et al. 2021(V21)\cite{vasiliev_tango_2021}) stems from the number of pericentric passages. As illustrated in the orbital evolution track (left panel of Supplementary Figure \ref{fig_massloss}), our simulation includes one additional pericentric passage. During this first close encounter with the Milky Way, the progenitor undergoes dramatic tidal stripping, losing a significant fraction of its extended dark matter halo.

As shown in the mass evolution plot (right panel of Supplementary Figure \ref{fig_massloss}), the total bound mass of the system drops rapidly after its first orbital period. In our analysis, the bound mass is rigorously defined kinematically in the Sgr rest frame: it includes only particles with negative total energy ($E < 0$). We impose a strict tracking condition: once a particle becomes unbound, it is permanently flagged as escaped and will not be re-classified as bound at later times, even if its trajectory temporarily brings it back to the progenitor's vicinity. Following this first orbit, the bound mass of our simulation drops to approximately $\sim 5\times 10^9\, M_\odot$. This post-stripping mass is comparable to the initial mass of $\sim 3.25\times 10^9\, M_\odot$ adopted by \cite{vasiliev_tango_2021}, which was spatially defined as the mass enclosed within $5\, {\rm kpc}$. The choice of a higher initial mass combined with a longer integration time robustly captures the early tidal stripping phase without qualitatively altering the chemo-kinematic signature of the resulting stream.

\subsection{The Impact of the Azimuthal Angle $\phi$ on the Stream}

To investigate the impact of the azimuthal angle ($\phi$) on the stream, we performed a suite of N-body simulations with the inclination fixed at our fiducial value ($\theta = 150^\circ$) while varying $\phi$. We first examined the effect of $\phi$ on the distribution of the stream.

Supplementary Figure~\ref{fig_xz_diffphi} presents the present-day distribution of the stream in the Galactocentric $X-Z$ plane, with the color coding indicating the metallicity. We observe that the overall structural distributions are highly similar across these models, and most configurations exhibit a vertical metallicity gradient in the regions corresponding to the observed leading and trailing arms.

Based on the positions of stellar particles, we then compared the trajectories of the stream. Supplementary Figure~\ref{fig_xz_traj_diffphi} displays the simulated stream tracks in the $X-Z$ plane for different $\phi$ values. As expected from the similar distributions, these tracks are highly coincident, showing almost no variation. In contrast, Supplementary Figure~\ref{fig_lam_beta_diffphi} compares the mean stream tracks in the Sgr coordinate frame ($\rm \widetilde{\Lambda}_\odot, \widetilde{B}_\odot$). Here, varying $\phi$ introduces transverse shifts among the models. This comparison clearly demonstrates that the impact of $\phi$ primarily manifests in the stream's on-sky trajectory, whereas its influence on the trajectory in the $X-Z$ plane is minimal.



\section{Bifurcation}

To investigate whether the observed ${\rm[Fe/H]}-L_y$ correlation arises from the superposition of the chemically distinct bright and faint branches, we performed a component-wise decomposition of the stream. We adopted the spatial classification criteria of Ramos et al. (2022)\cite{ramos_2022} to separate the sample into four distinct components: leading-bright, leading-faint, trailing-bright, and trailing-faint branches based on their distribution in the Sgr coordinate frame ($\rm \widetilde{\Lambda}_\odot, \widetilde{B}_\odot$)(Supplementary Figure \ref{lambda_beta}). 

We evaluated the chemo-kinematic signature for each component individually. As shown in Supplementary Figure \ref{ly_fe_bifurcation}, the characteristic ${\rm[Fe/H]}-L_y$ correlation is recovered within both the bright and faint branches of the leading and trailing arms. The persistence of this signal across individual branches confirms that the correlation is an intrinsic property of the Sgr stream.

We further compare our fiducial model ($\theta={150}^\circ,\phi={40}^\circ$) with the observational data from Ramos et al. (2022)\cite{ramos_2022} in the Supplementary Figure \ref{fig_compare_ramos}. The simulated stream tracks show good agreement with the observed global trajectory (top panel) and the kinematic trends along the stream (bottom panels). Our fiducial model is optimized to successfully capture the ${\rm[Fe/H]}-L_y$ correlation, but it does not simultaneously reproduce the prominent spatial bifurcation seen in the data.

Indeed, previous dynamical studies have established that the stream bifurcation is highly sensitive to the initial disk orientation (defined by angles $\theta$ and $\phi$)\cite{penarrubia_was_2010,oria_revisiting_2022}. Our grid exploration reveals a trade-off between reproducing the bifurcation and the ${\rm[Fe/H]}-L_y$ correlation in both leading arm and trailing arm. For instance, while configurations like the model shown in Supplementary Figure \ref{fig_sim_bifurcation} ($\theta=150^\circ,\phi=150^\circ$) do successfully reproduce both the bifurcation and the ${\rm[Fe/H]}-L_y$ correlation in the trailing arm, they simultaneously weaken the correlation in the leading arm.

This trade-off suggests that generating both features simultaneously requires more advanced dynamical frameworks. Consequently, the existence of this vertical metallicity gradient emerges as a constraint for future theoretical modeling. Any successful mechanism proposed to explain the bifurcation should not only reproduce the spatial split but also naturally preserve the ${\rm[Fe/H]}-L_y$ correlation. Testing this chemo-kinematic signature in future spectroscopic surveys will be paramount for uncovering the true origin of the bifurcation.

\section{Sagittarius core}



Recent studies have revealed a complex metallicity distribution within the Sgr core. Hayes et al. (2020)\cite{Hayes_2020} and Minelli et al. (2023)\cite{Minelli_2023} found that the core is dominated by metal-rich stars. However, studies using the Pristine survey also identified a significant metal-poor population \cite{vitali_2022, Sestito_2024}. This suggests that the Sgr progenitor likely contained multiple stellar populations rather than a single metallicity component.

In Supplementary Figure \ref{fig_meta_core} (Left), we plot the metallicity distribution as a function of angular distance from the Sgr center, combining data from Hayes et al. (2020)\cite{Hayes_2020} and Sestito et al. (2024)\cite{Sestito_2024}. It is crucial to note that these surveys possess distinct and strong selection functions. Therefore, their direct, uncorrected combination does not represent the true intrinsic Metallicity Distribution Function (MDF) of the remnant, which is likely asymmetric or radially varying rather than strictly bimodal. Nevertheless, as a qualitative toy model to capture the coexistence of a metal-poor component and a distinct metal-rich population (centered at ${\rm[Fe/H]}\sim-0.5$) in our initial conditions, we refine $comp.1$ by replacing the original single Gaussian distribution with a double Gaussian profile, as shown in Supplementary Figure \ref{fig_meta_core} (Middle).

Supplementary Figure \ref{fig_meta_core} (Right) presents the outcome of this simulation. While this empirical setup broadly accommodates the wide metallicity spread observed in the remnant core, we emphasize that it is not intended to be a definitive physical reconstruction of the core's true chemical evolution. Furthermore, the simulation reveals that tidal forces induce significant radial migration: stars currently observed in the core region originally resided in the deepest parts of the progenitor's potential well. Crucially, we find that modifying the chemical composition of the extreme inner core has a negligible impact on the ${\rm[Fe/H]}-L_y$ correlation in the extended stream (Supplementary Figure \ref{fig_core_gradient}). This reinforces that the stream's chemo-kinematic properties are driven by the region($1.5 - 11.5$ kpc), largely independent of the core's detailed complexity.

Previous study has utilized the present-day angular momentum of the Sgr core to constrain the initial orientation of the progenitor disk\cite{Pino_2021}. However, the current kinematics of the remnant core may not be a definitive tracer of the progenitor. Simulation studies by \cite{penarrubia_no_2011,frinchaboy_2mass_2012,vasiliev_last_2020} have demonstrated that the pressure-supported (spheroidal) progenitors can also successfully reproduce the observed morphology and kinematics of the Sgr core.

Our fiducial model, optimized to reproduce the chemo-kinematics of the stream, predicts a core rotation that differs from the observed Sgr core. The discrepancy likely points to a more complex internal structure for the Sgr progenitor than a simple single-component disk. A plausible scenario involves a hybrid system: a pressure-supported core embedded within a rotation-supported disk \cite{oria_revisiting_2022}. While exploring such multi-component models is beyond the scope of this work, our results underscore the need to integrate both stream chemistry and core dynamics in future reconstructions of the Sagittarius dwarf galaxy.

\section{Impact of Large Magellanic Cloud}

In our MW+LMC setup, the adoption of a spherical dark matter halo for the MW naturally results in the simulated stream not being perfectly aligned with the observed one. Because the mock catalog is extracted strictly within the observed spatial coverage, this spatial offset—which is particularly pronounced for the leading arm—leads to incomplete sampling at the simulated stream's edges (e.g., missing the metal-poor stars on the outer flank of the misaligned leading arm). Consequently, a global view of the mock sample (Supplementary Figure \ref{supp_fig8}) may visually weaken the underlying ${\rm[Fe/H]}-L_y$ correlation.

As established earlier using the observational sample (see Supplementary Figure \ref{fig_lambda_bin}), the ${\rm[Fe/H]}-L_y$ correlation is fundamentally a local feature present within individual $\tilde{\Lambda}_\odot$ bins. To overcome the sampling artifact introduced by the spatial misalignment, the simulated stream should also be evaluated locally (see Supplementary Figure \ref{fig_sim_bin}). We specifically focus on the longitudinal ranges of $105^\circ < \tilde{\Lambda}_\odot < 125^\circ$ for the leading arm and $-120^\circ < \tilde{\Lambda}_\odot < -100^\circ$ for the trailing arm, as these represent the segments where the observed ${\rm [Fe/H]}-L_y$ correlation is most significant. When the simulated stream is examined independently within these regions, the 
${\rm [Fe/H]}-L_y$ correlation is robustly preserved despite the LMC's perturbation.

These results demonstrate that the intrinsic physical link between the stream's present-day vertical metallicity gradient and the progenitor's metallicity profile survives external time-dependent perturbations. While the MW+LMC calculation serves as a representative proof-of-concept, it strengthens the scenario that the observed gradient originates from an initial $\wedge$-like metallicity profile.

\section{Comparison of disk model, spheroidal model, and rotating elliptical model }
 
In this section, we compare the stellar streams generated by the fiducial model and the other two galaxies. For the spheroidal model and rotating elliptical model, it is not a preferable option to mark the metallicity of each particle according to $ l_z$ like the equation 4 of Method. In the spheroidal model, we use $r$ in place of $l_z$. In the rotating elliptical model, we employ
$
\widetilde{r} = \sqrt{R^2 + (a/c)^2 z^2}
$
as the analogous variable. These three models all have the $\wedge$-like metallicity profile.

To investigate the causes of the  ${\rm[Fe/H]}-L_y$ correlation, we analyze all the particles of three models within the observation area and do not add errors to the simulated data. Supplementary Figure \ref{strip} presents the distributions of leading arms and trailing arms in the ${\rm[Fe/H]}-L_y$ space produced by various models, and the color of each point represents the average stripped time of the particles in this bin. 


As illustrated by the disk model in Supplementary Figure \ref{strip}, the ${\rm[Fe/H]}-L_y$ correlation arises naturally from stellar particles stripped at different epochs. In this scenario, particles with distinct stripping times retain distinct angular momenta and metallicities. Conversely, this correlation is absent in both the spheroidal and rotating elliptical models. In those two scenarios, the progenitor's low rotational velocity and large velocity dispersion completely wash out the distinct angular momentum signatures. 

\makeatletter
\renewcommand{\fnum@table}{\textbf{Supplementary Table~\thetable}}
\makeatother

\begin{table}[h]
\caption{\textbf{Structural and numerical parameters of the fiducial model.} This table presents the initial configuration of the simulated Sagittarius progenitor set up at 3.6 Gyr ago. For the dark matter (DM) halo modeled with a Hernquist profile, $M_{200}$ denotes the virial mass in units of $10^{10}M_{\odot}$, and $N_{\rm dm}$ and $\epsilon_{\rm dm}$ represent the particle number and gravitational softening length, respectively. For the exponential stellar disk, the numerical parameters include the particle numbers for the two components ($N_{\rm disk}\ \text{comp.1}$ and $\text{comp.2}$) and the disk softening length $\epsilon_{\rm disk}$. The disk orientation is specified by the spherical coordinates $\theta$ and $\phi$.}\label{tab1}%
\begin{tabular}{@{}llll@{}}
\toprule
Component &  Parameter & value \\
\midrule
DM halo      &  mass,$M_{200}$($10^{10}M_{\odot}$)  & 2   \\
(Hernquist)  &  concentration                       & 10  \\
             &  $N_{\rm dm}$                            & 200,000 \\
             &  $\epsilon_{\rm dm}$(kpc)                & 0.1 \\
\midrule
Disk         &  mass($10^8M_{\odot}$)               & 8   \\
(Exponential)&  scale length($\rm kpc$)                 & 2.5  \\
             &  scale height($\rm kpc$)                 & 0.75 \\
             &  orientation $\theta$($^\circ$)      & 150 \\         
             &  orientation $\phi$($^\circ$)        & 40 \\   
             &  $N_{\rm disk}\ (comp.1)$           & $\sim 46,000$ \\
             &  $N_{\rm disk}\ (comp.2)$            & $\sim 154,000$ \\
             &  $\epsilon_{\rm disk}$(kpc)                & 0.02 \\
\botrule
\end{tabular}
\end{table}

\begin{table}[h]
\caption{\textbf{Initial configurations of alternative progenitor models.} This table compares the fundamental structural, kinematic, and numerical parameters between the control spheroidal model and the rotating elliptical model. Stellar mass is given in units of $M_{\odot}$, and $a$ specifies the scale length in $\rm kpc$. The parameter $a/c$ indicates the stretch or flattening ratio of the components. The kinematic spinning factor is represented by $k$, while $N(\text{comp.1})$ and $N(\text{comp.2})$ denote the stellar particle numbers allocated for the two respective components within each model.}\label{tab2}%
\begin{tabular}{@{}lll@{}}
\toprule
Parameter & Spheroidal & Rotating elliptical \\
\midrule
Stellar Mass            & $8\times10^{8}M_{\odot}$ & $8\times10^{8}M_{\odot}$ \\
Scale length ($a$)      & $1.53\ \rm kpc$          & $2.29\ \rm kpc$          \\
Stretch ($a/c$)         & 1                        & 2                        \\
$k$                     & 0                        & 1                        \\
$N({\rm comp.1})$       & $\sim 52,000$            & $\sim 42,000$            \\
$N({\rm comp.2})$       & $\sim 148,000$           & $\sim 158,000$           \\
\botrule
\end{tabular}
\end{table}

\begin{figure}[h]
    \centering
    \includegraphics[width=10cm,height=7.5cm]{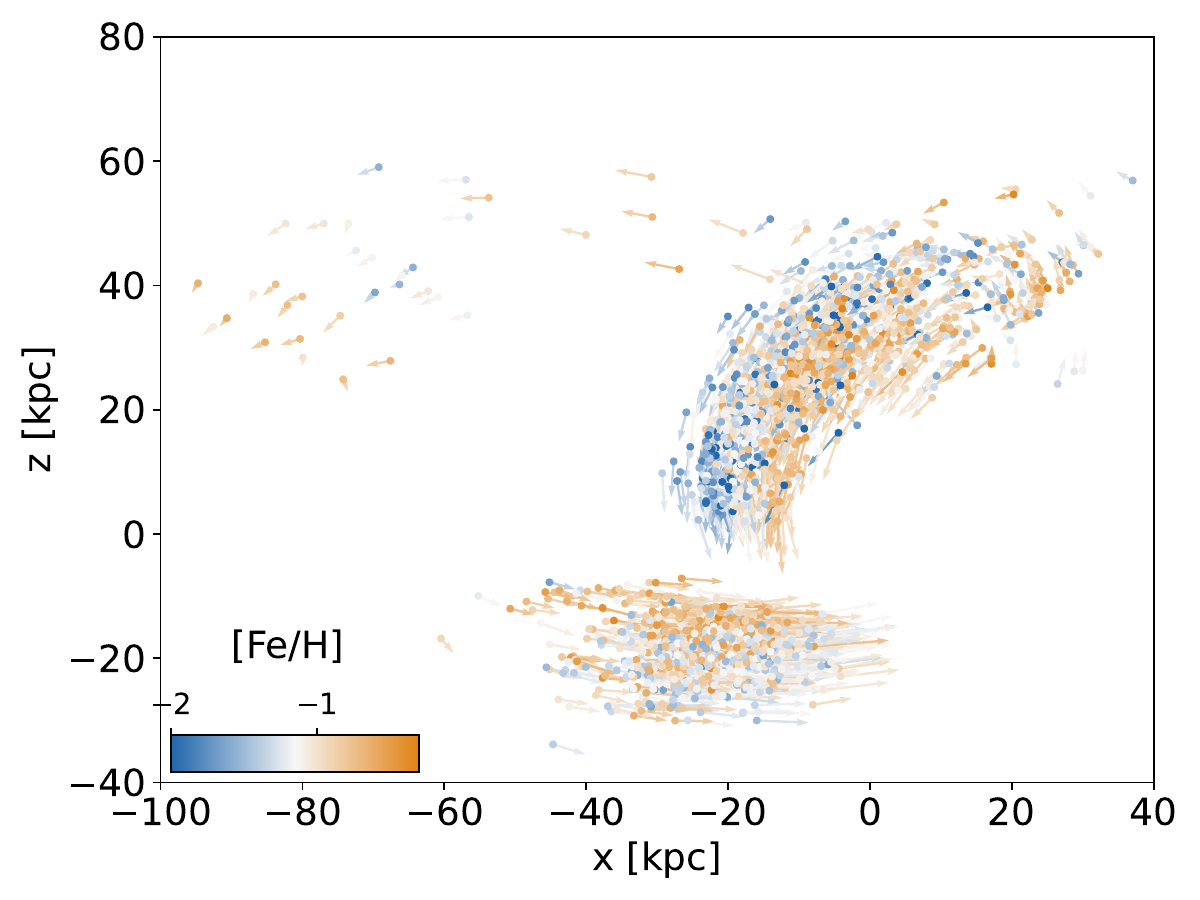}
    \caption{\textbf{Distribution of the mock data. }The spatial distribution of the particles of the simulation in the $X-Z$ plane. Arrows represent the stellar particles’ moving directions and velocity amplitudes, and each particle is color-coded according to its metallicity.} 
    \label{sim_pos}
\end{figure}

\begin{figure}
    \centering
    \includegraphics[width=15cm,height=15cm]{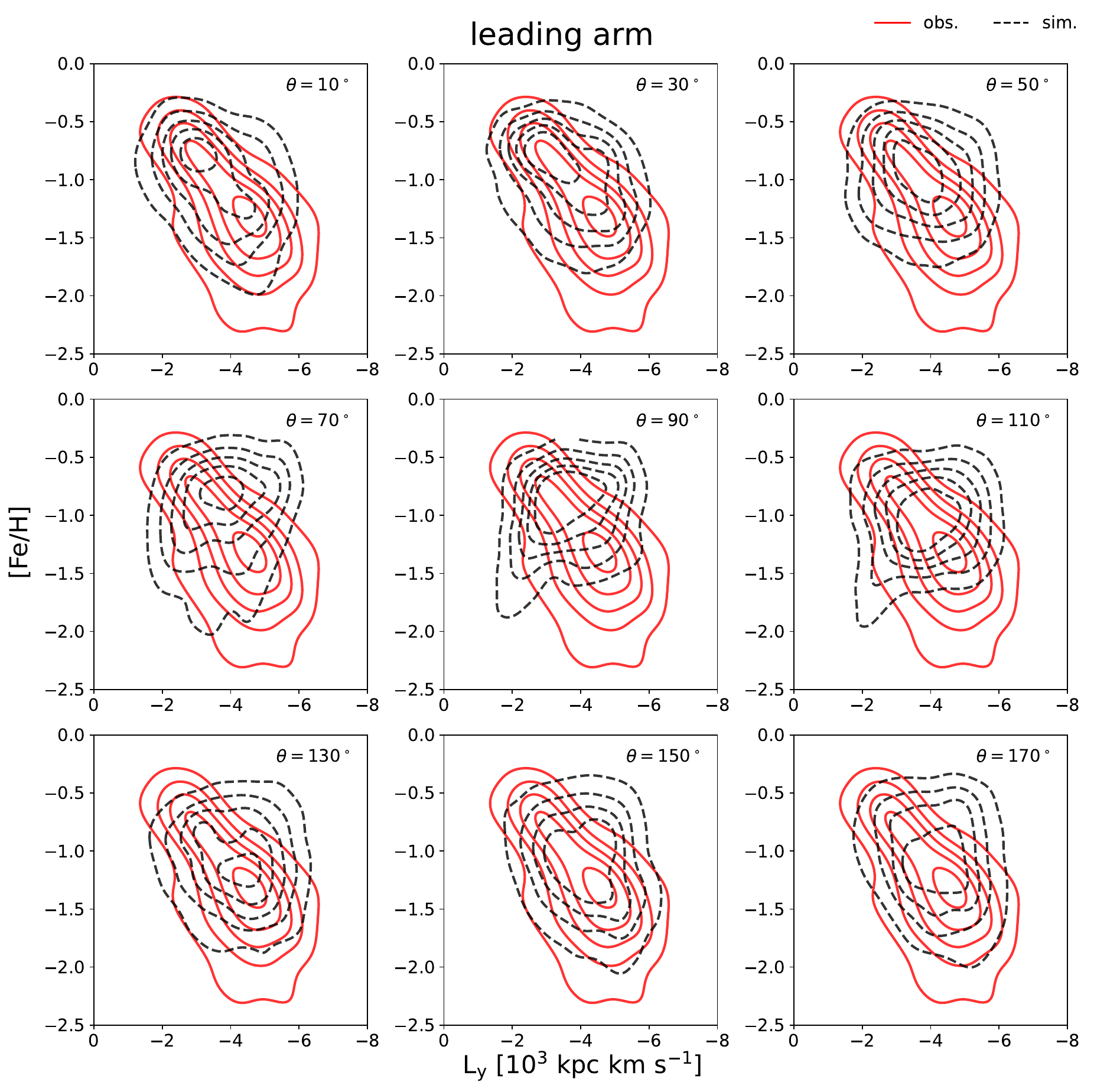}
    \caption{\textbf{The impact of different $\theta$ on leading arm in ${\rm[Fe/H]}-L_y$ plane. }In these simulations, $\phi$ is taken as 0. The red contour represents the density of observation points, while the gray contour shows the distribution of points sampled from the simulation data.
    }
    \label{theta_leading}
\end{figure}

\begin{figure}
    \centering
    \includegraphics[width=15cm,height=15cm]{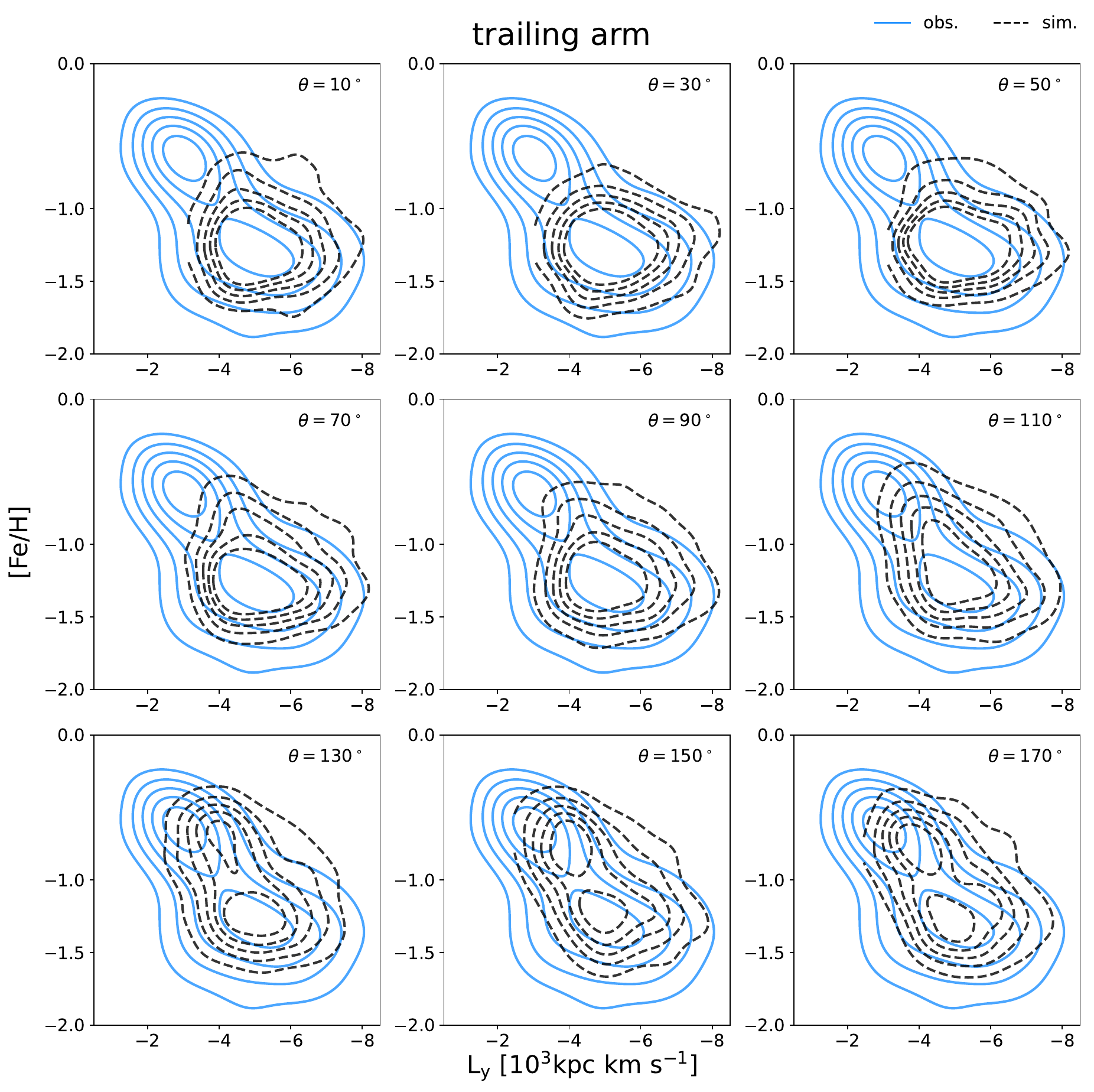}
    \caption{\textbf{The impact of different $\theta$ on trailing arm in ${\rm[Fe/H]}-L_y$ plane. }In these simulations, $\phi$ is taken as 0. The blue contour represents the density of observation points, while the gray contour shows the distribution of points sampled from the simulation data.}
    \label{theta_trailing}
\end{figure}

\begin{figure}
    \centering
    \includegraphics[width=14cm,height=7cm]{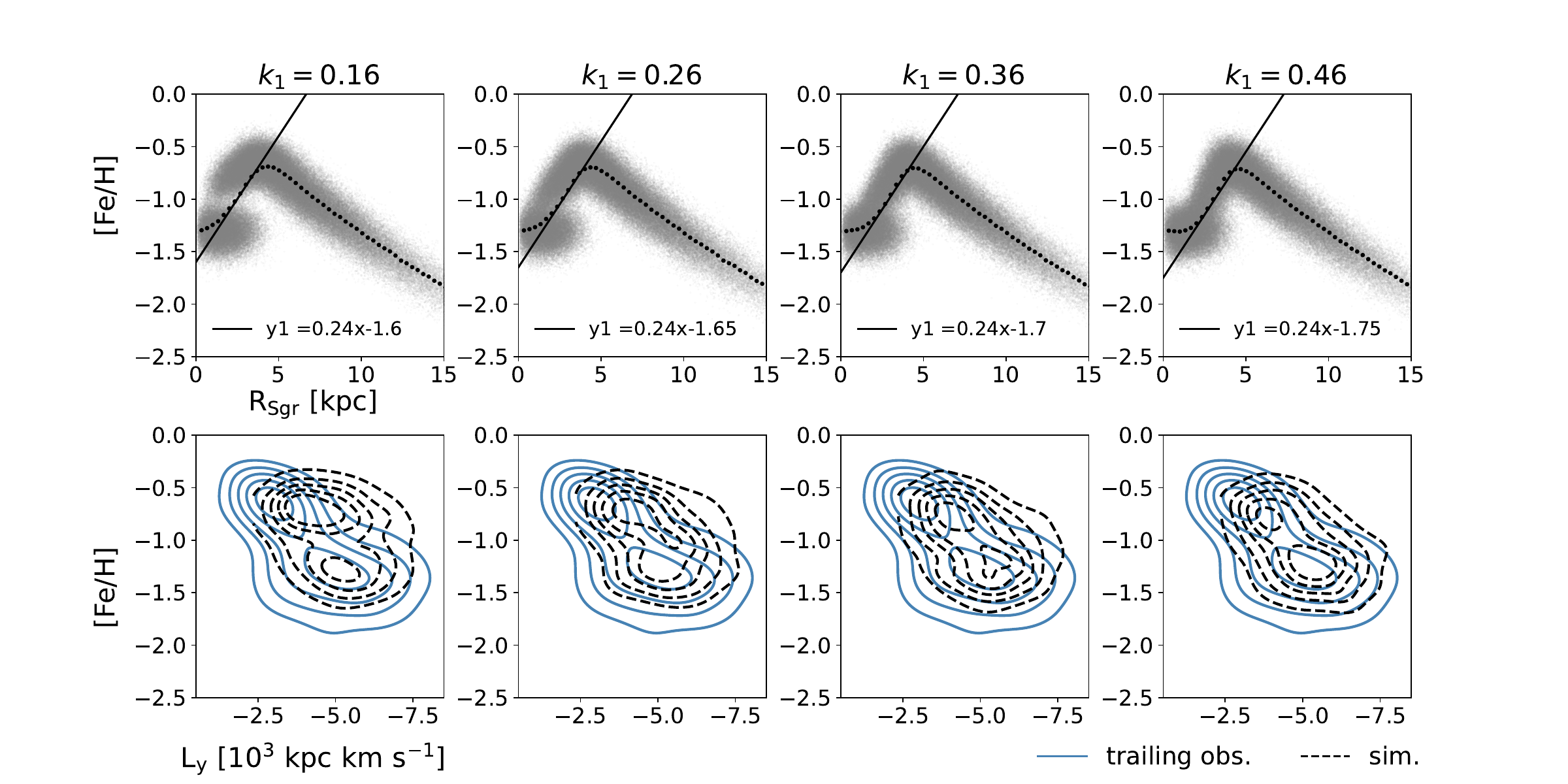}
    \caption{\textbf{The impact of $k_1$ in ${\rm[Fe/H]}-L_y$ plane.} \textbf{Top row:} Initial metallicity (${\rm [Fe/H]}$) profiles of the Sgr disk as a function of radius $\rm R_{Sgr}$ under various $k_1$ values, with $k_2$ fixed at 0.22. In each panel, the background gray dots represent individual stellar particles, displaying their initial distances from the center of Sgr and their corresponding metallicities. The black points track the mean metallicity calculated at each radial bin. The solid line represents the initial metallicity gradient of the rising branch derived from these mean metallicities. \textbf{Bottom row:} Distributions in the ${\rm [Fe/H]} - L_y$ plane illustrating the impact of different $k_1$ values. The blue contours represent the observed ${\rm [Fe/H]} - L_y$ correlation of the trailing arm, while the black dashed contours depict the simulated distribution from the respective models.}\label{ed_fig4_5}
    \label{k1_slope}
\end{figure}

\begin{figure}
    \centering
    \includegraphics[width=14cm,height=7cm]{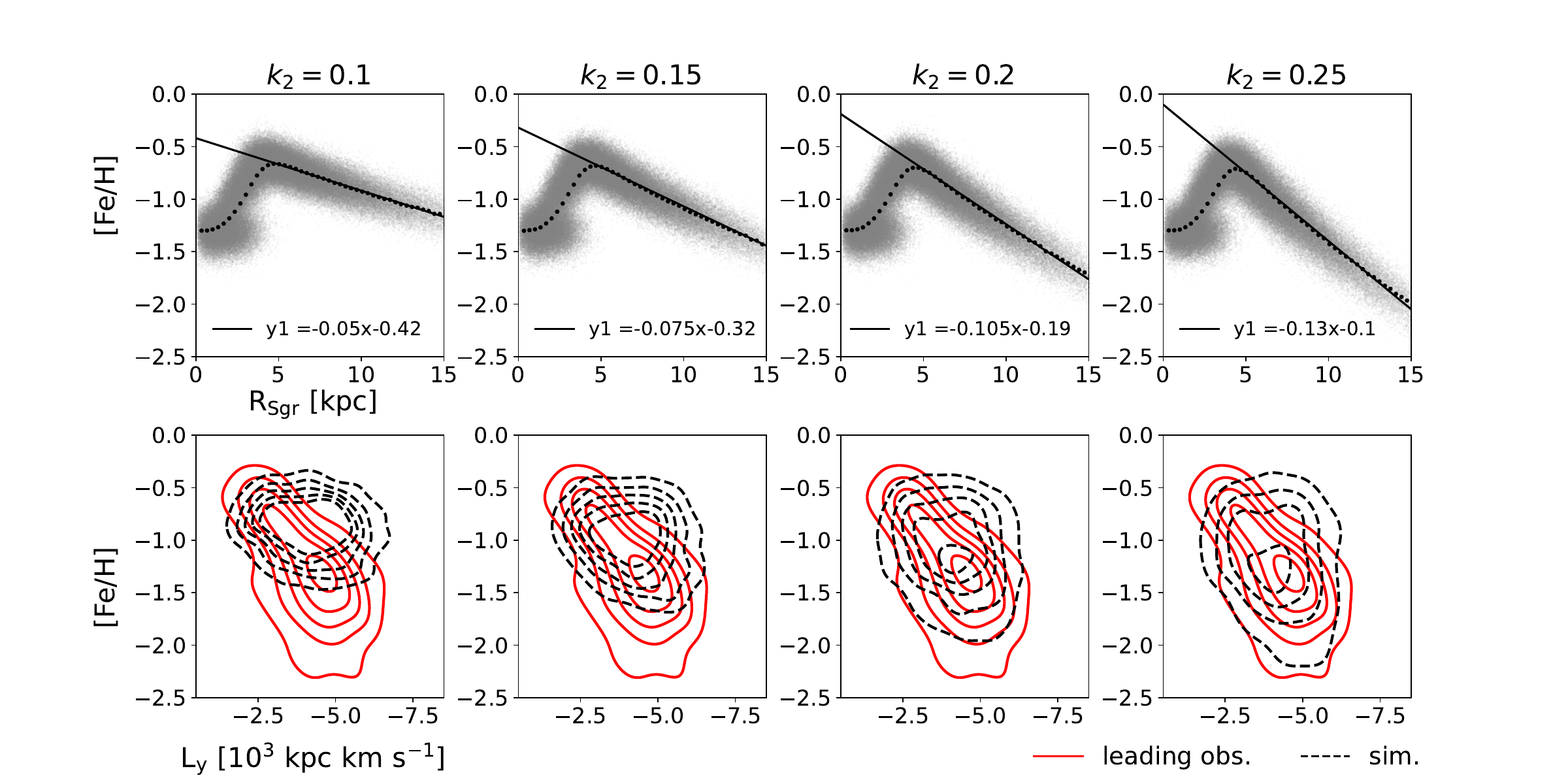}
    \caption{\textbf{The impact of $k_2$ in ${\rm[Fe/H]}-L_y$ plane.} \textbf{Top row:} Initial metallicity (${\rm [Fe/H]}$) profiles of the Sgr disk as a function of radius $\rm R_{Sgr}$ under various $k_2$ values, with $k_1$ fixed at 0.36. In each panel, the background gray dots represent individual stellar particles, displaying their initial distances from the center of Sgr and their corresponding metallicities. The black points track the mean metallicity calculated at each radial bin. The solid line represents the initial metallicity gradient of the rising branch derived from these mean metallicities. \textbf{Bottom row:} Distributions in the ${\rm [Fe/H]} - L_y$ plane illustrating the impact of different $k_2$ values. The red contours represent the observed ${\rm [Fe/H]} - L_y$ correlation of the leading arm, while the black dashed contours depict the simulated distribution from the respective models.}
    \label{k2_slope}
\end{figure}

\begin{figure}
    \centering
    \includegraphics[width=14cm,height=7cm]{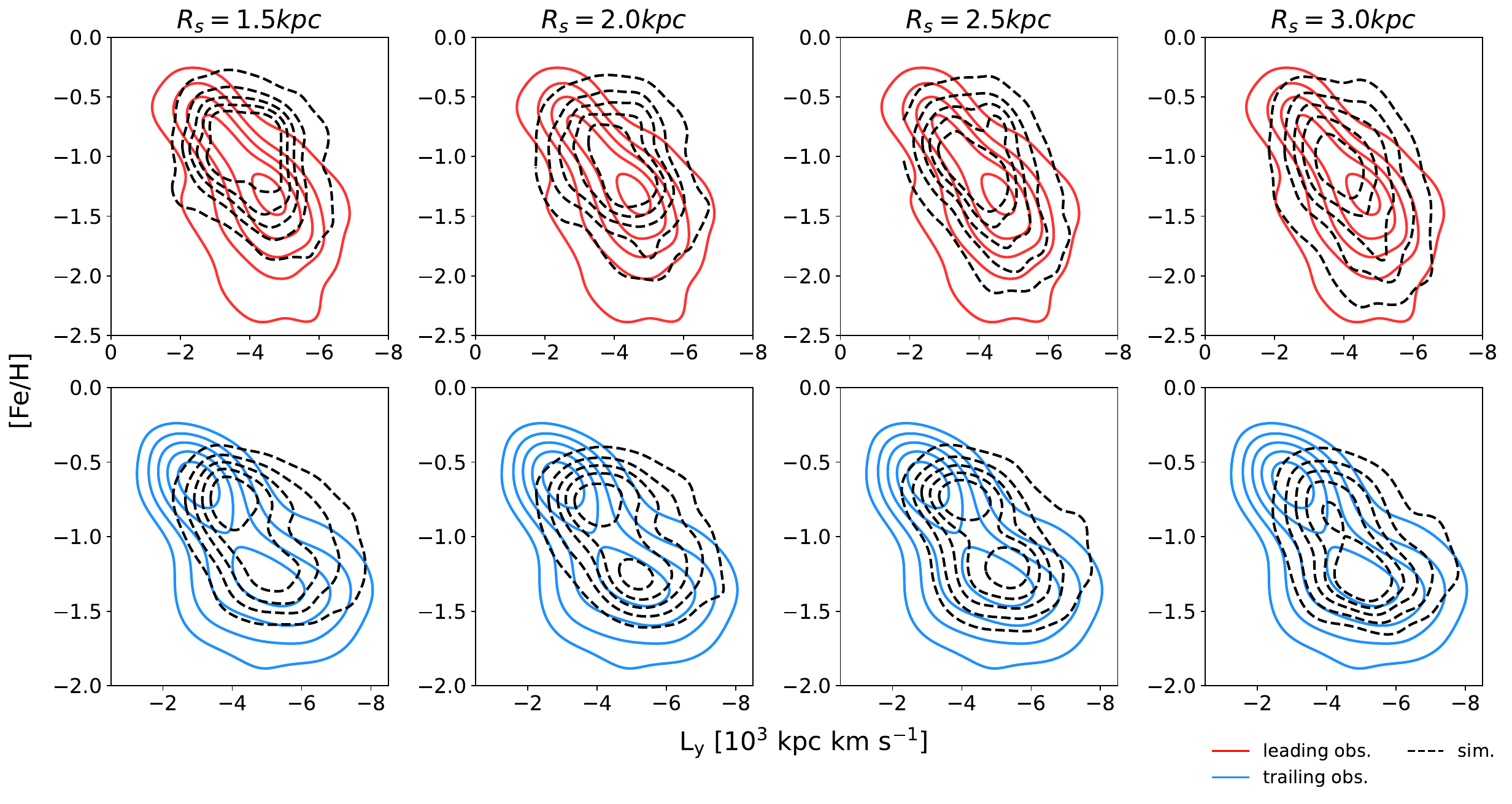}
    \caption{\textbf{The influence of scale length on the stream in the ${\rm[Fe/H]}-L_y$ plane.} The parameter $R_{\rm s}$ denotes the initial scale length of the Sgr dwarf galaxy model. \textbf{Top row:} Distributions of the simulated leading arm under different $R_{\rm s}$. The red contours represent the observed ${\rm [Fe/H]} - L_y$ correlation of the leading arm. \textbf{Bottom row:} Distributions of the simulated trailing arm under different $R_{\rm s}$. The blue contours represent the observed distribution of the trailing arm. In all panels, the simulated mock data from the models are depicted as black dashed contours overplotted on the observational data for direct comparison.}\label{ed_fig6}
\end{figure}

\begin{figure}
    \centering
    \includegraphics[width=14cm,height=7cm]{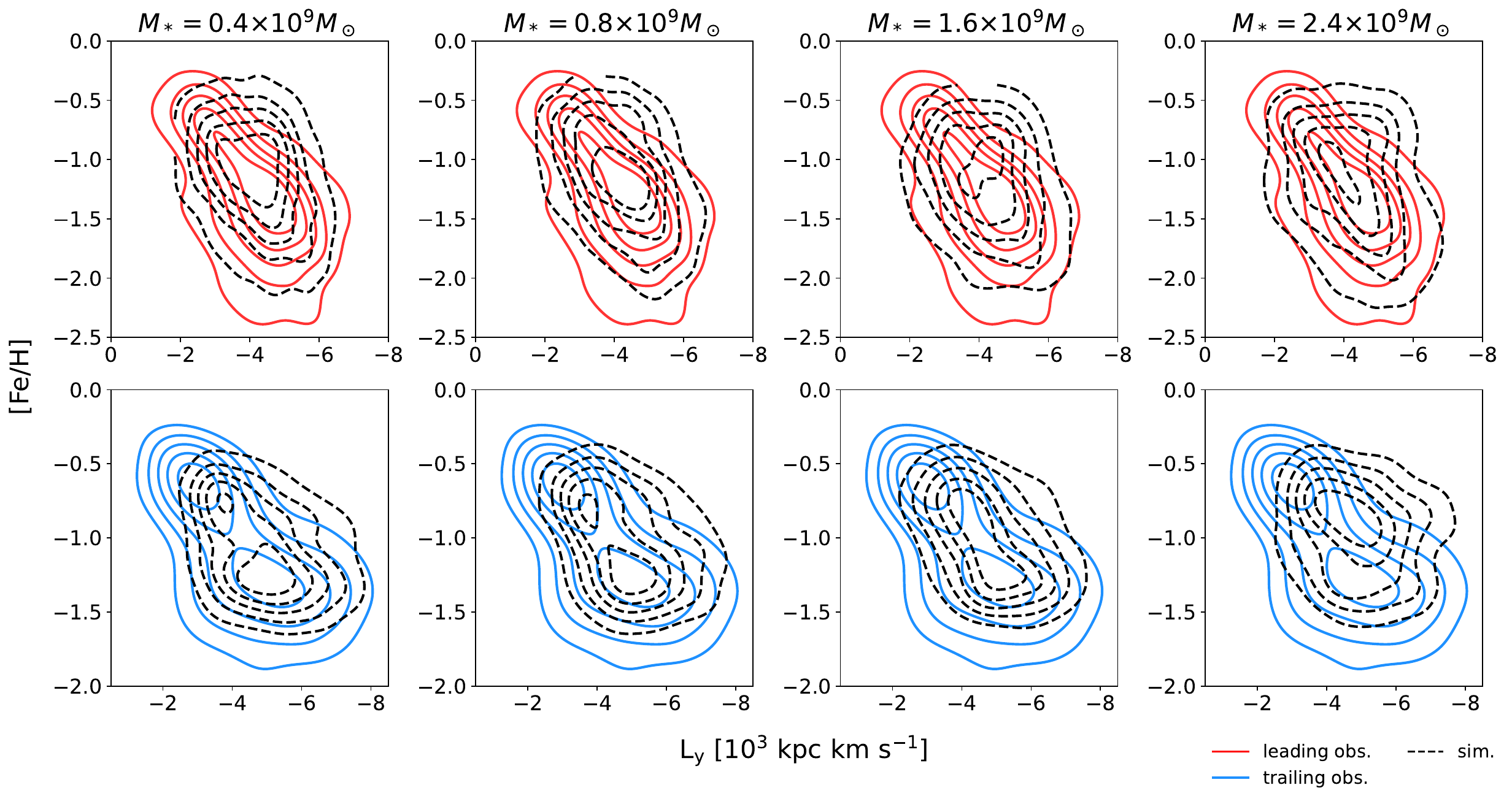}
    \caption{\textbf{The influence of stellar mass on the stream in the ${\rm[Fe/H]}-L_y$ plane.} 
    The parameter $M_*$ denotes the initial stellar mass of the Sgr progenitor model. \textbf{Top row:} Distributions of the simulated leading arm under different $M_*$. The red contours represent the observed ${\rm [Fe/H]} - L_y$ correlation of the leading arm. \textbf{Bottom row:} Distributions of the simulated trailing arm under different $M_*$. The blue contours represent the observed distribution of the trailing arm. In all panels, the simulated mock data from the models are depicted as black dashed contours overplotted on the observational data for direct comparison.}
    \label{mass}
\end{figure}

\begin{figure}
    \centering
    \includegraphics[width=10cm,height=5cm]{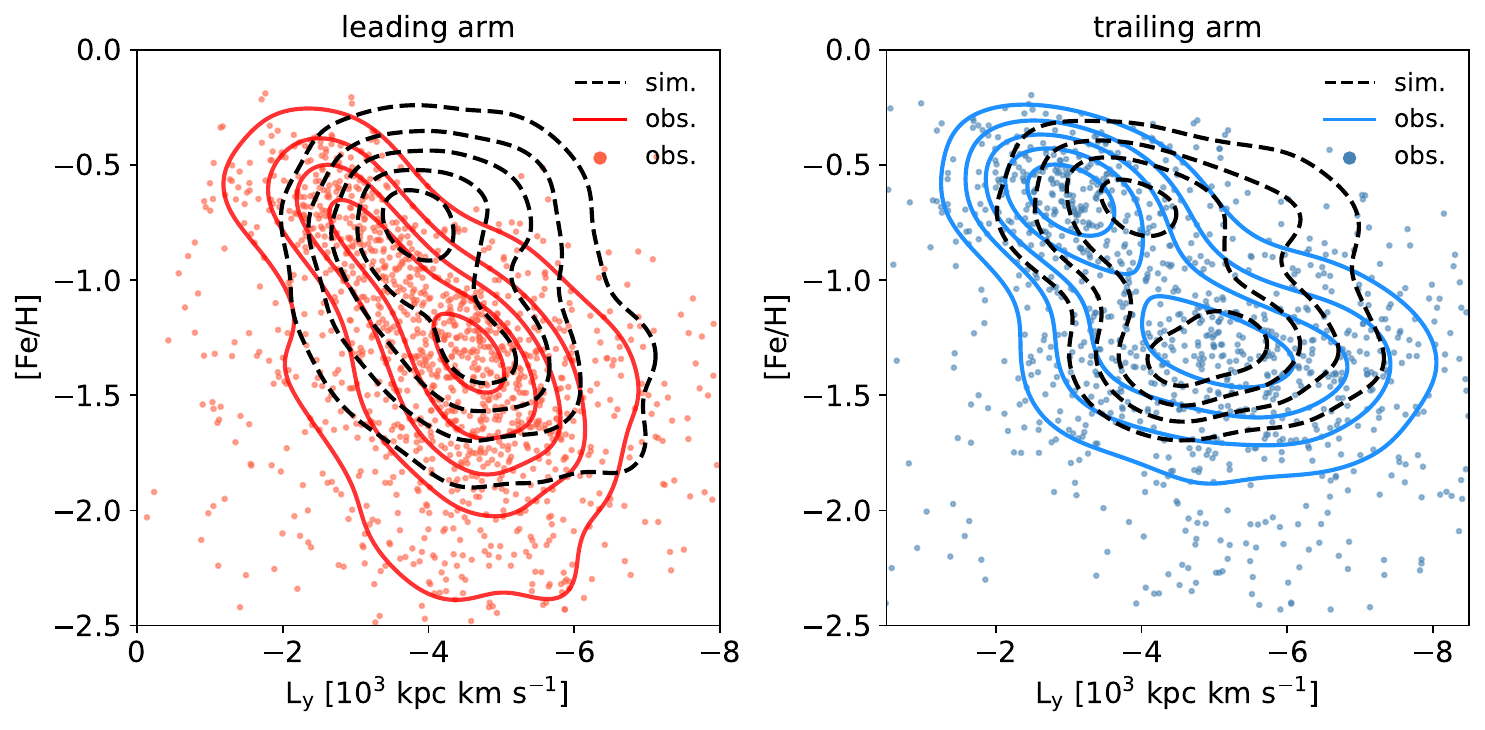}
    \caption{\textbf{The influence of the time-dependent potential on the ${\rm[Fe/H]}-L_y$ correlation}. The initial orientation of the progenitor Sgr model is configured with $\theta = 130^\circ$ and $\phi = 0^\circ$. These panels illustrate the ${\rm[Fe/H]} - L_y$ correlation evolved within the time-dependent Galactic potential including the Large Magellanic Cloud (MW+LMC). 
    \textbf{Left:} The red contour and red dots represent the observational data, while the black contour indicates the simulated ${\rm[Fe/H]} - L_y$ correlation of the leading arm. \textbf{Right:} The blue contour and blue dots represent the observational data, while the black contour indicates the simulated ${\rm[Fe/H]} - L_y$ correlation of the trailing arm.}\label{supp_fig8}
\end{figure}

\begin{figure}[h]
    \centering
    \includegraphics[width=10cm,height=7cm]{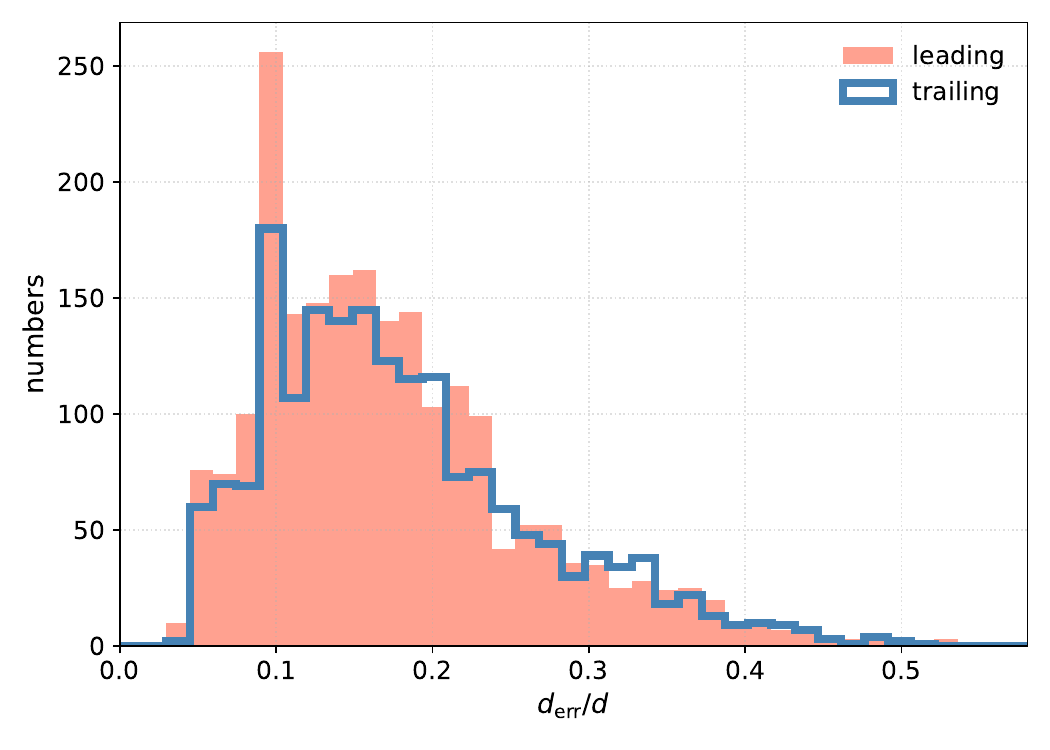}
    \caption{\textbf{The distribution of relative distance errors $(d_{\rm err}/d)$ for the Sagittarius stream members. The red histogram represents the leading arm, and the blue histogram represents the trailing arm. The majority of the sample possesses distance uncertainties below 20\%, supporting the reliability of our spatial and kinematic analysis.}}
    \label{fig_derr_bin}
\end{figure}

\begin{figure}[h]
    \centering
    \includegraphics[width=12cm,height=12cm]{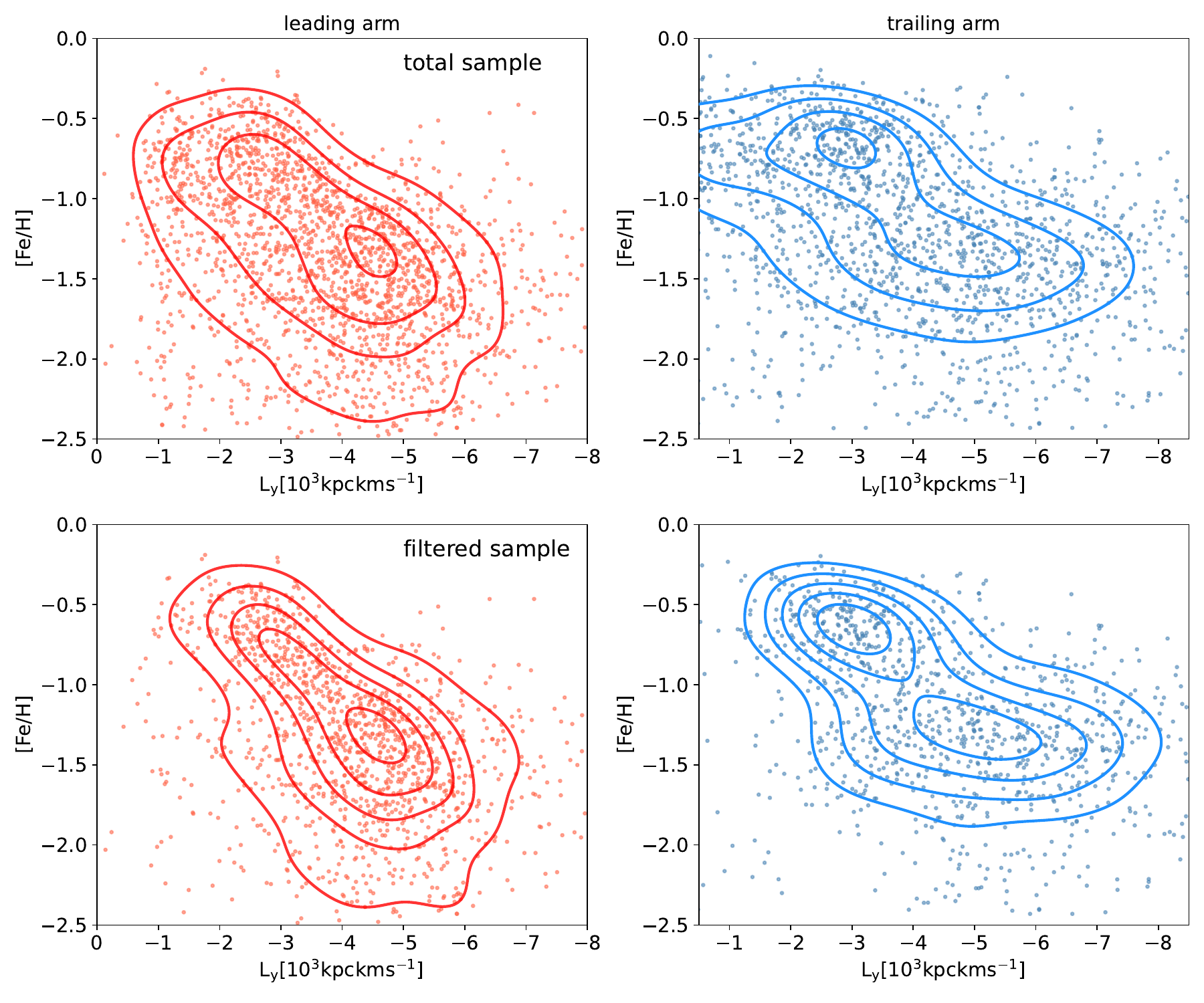}
    \caption{\textbf{Comparison of the ${\rm[Fe/H]}-L_y$ distribution before and after applying data quality cuts.}  \textbf{Top row:} The initial, uncut kinematic candidate sample. \textbf{Bottom row:} The sample of Sgr members after applying our fiducial quality cuts ($d_{\rm err}/d < 0.2$, $v_{\rm los, err} < 15\ {\rm km\,s^{-1}}$, and ${\rm [Fe/H]}_{\rm err} < 0.25\ {\rm dex}$). The persistence of the [Fe/H]-Ly correlation across both panels demonstrates that the observed trend is robust and not an artifact of our data selection.}
    \label{fig_derr_fe_ly}
\end{figure}

\begin{figure}[h]
    \centering
    \includegraphics[width=13cm,height=4.5cm]{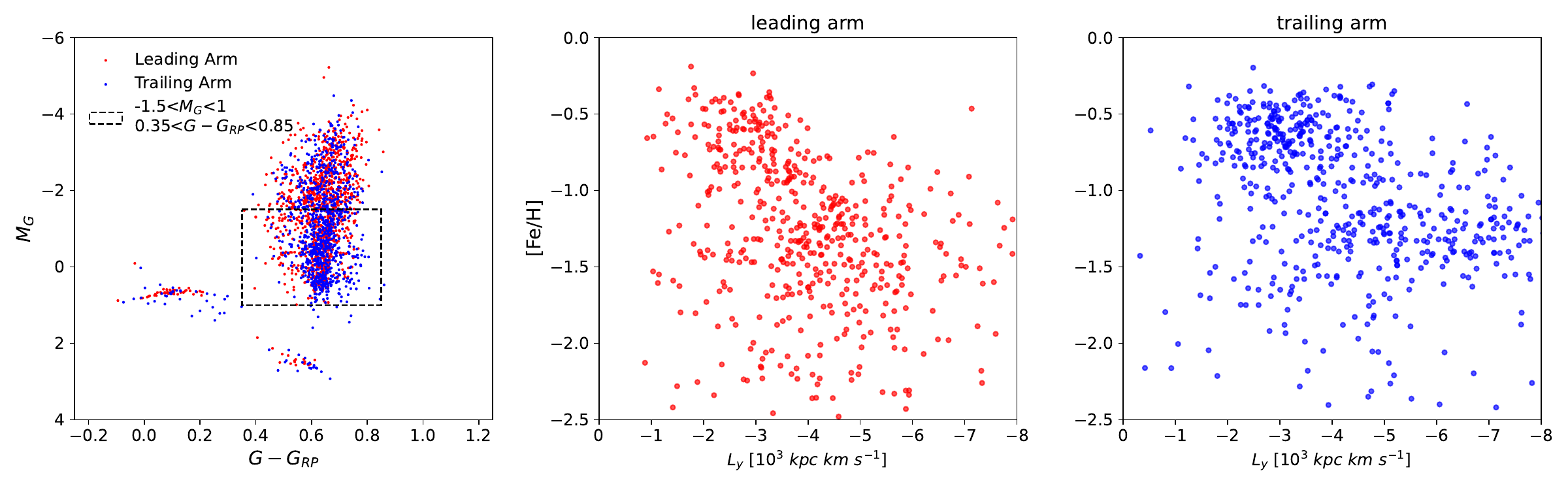}
    \caption{\textbf{ Robustness check using a restricted color-magnitude window.  Left panel: The Color-Magnitude Diagram (CMD) of the stream sample. Red and blue points correspond to stars in the Leading and Trailing arms, respectively. The black dashed rectangle indicates the uniform selection window defined by  $-1.5 < M_G < 1.0$ and $0.35 < (G-G_{\rm RP}) < 0.85$. Middle and Right 
    panels: The ${\rm[Fe/H]}-L_y$ correlation for the stars selected within the uniform window (i.e., inside the dashed box). 
    The middle panel displays the leading arm, and the right panel displays the trailing arm. The correlation remains evident even in this restricted subsample.}}
    \label{fig_mag_cut}
\end{figure}

\begin{figure}[h]
    \centering
    \includegraphics[width=13cm,height=7cm]{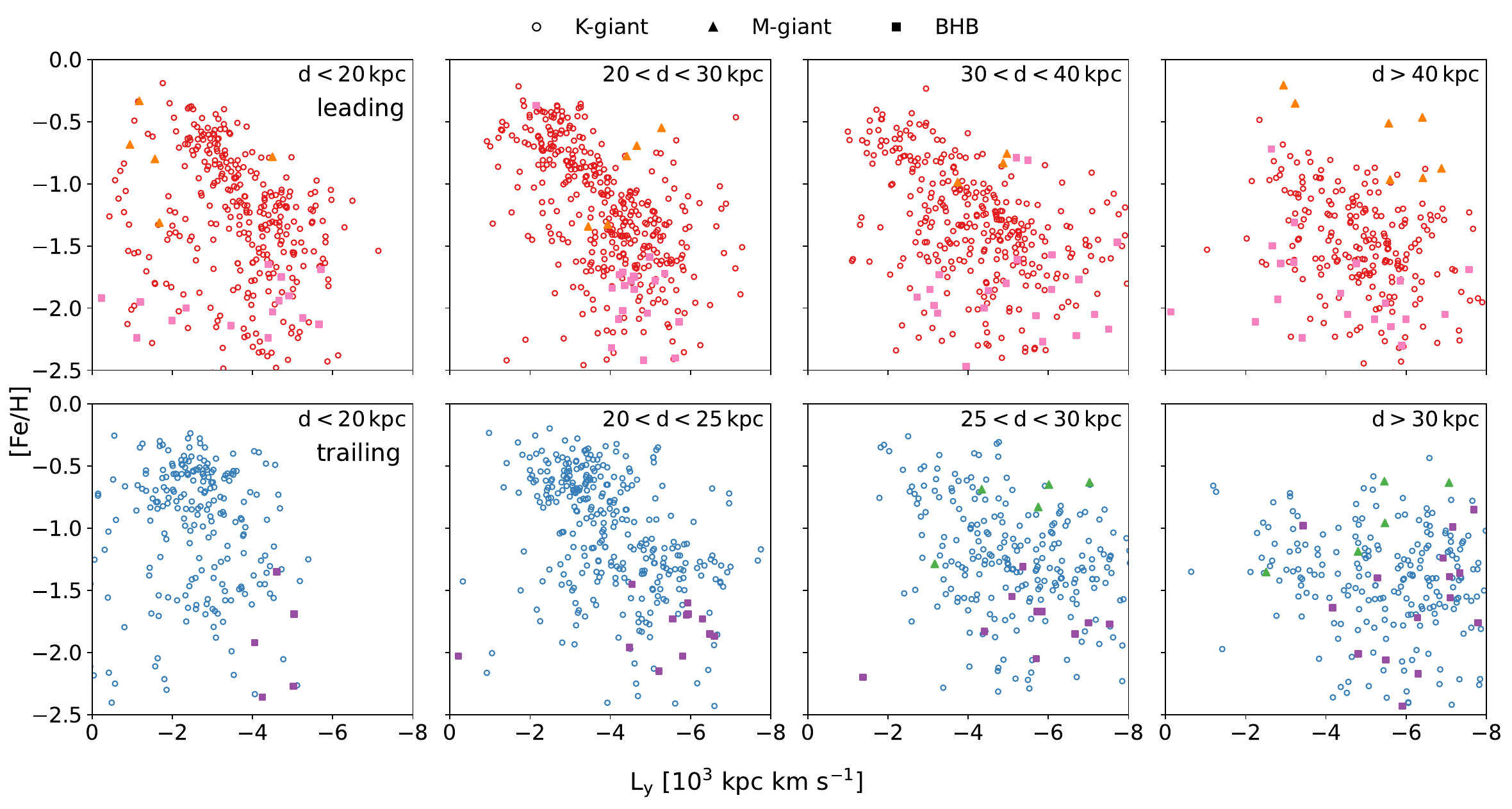}
    \caption{\textbf{ Robustness of the ${\rm[Fe/H]}-L_y$ correlation against distance and tracer effects.} The sample is divided into heliocentric distance bins of 10 kpc width for the leading arm (top row) and trailing arm (bottom row). Symbols distinguish different tracer populations: circles (K-giants), triangles (M-giants), and squares (BHBs). The characteristic correlation persists across all distance bins for the leading arm. For the trailing arm, the trend remains visible, particularly in the 20-25 kpc and 25-30 kpc ranges.}
    \label{tracer_distance}
\end{figure}

\begin{figure}[h]
    \centering
    \includegraphics[width=13cm,height=6cm]{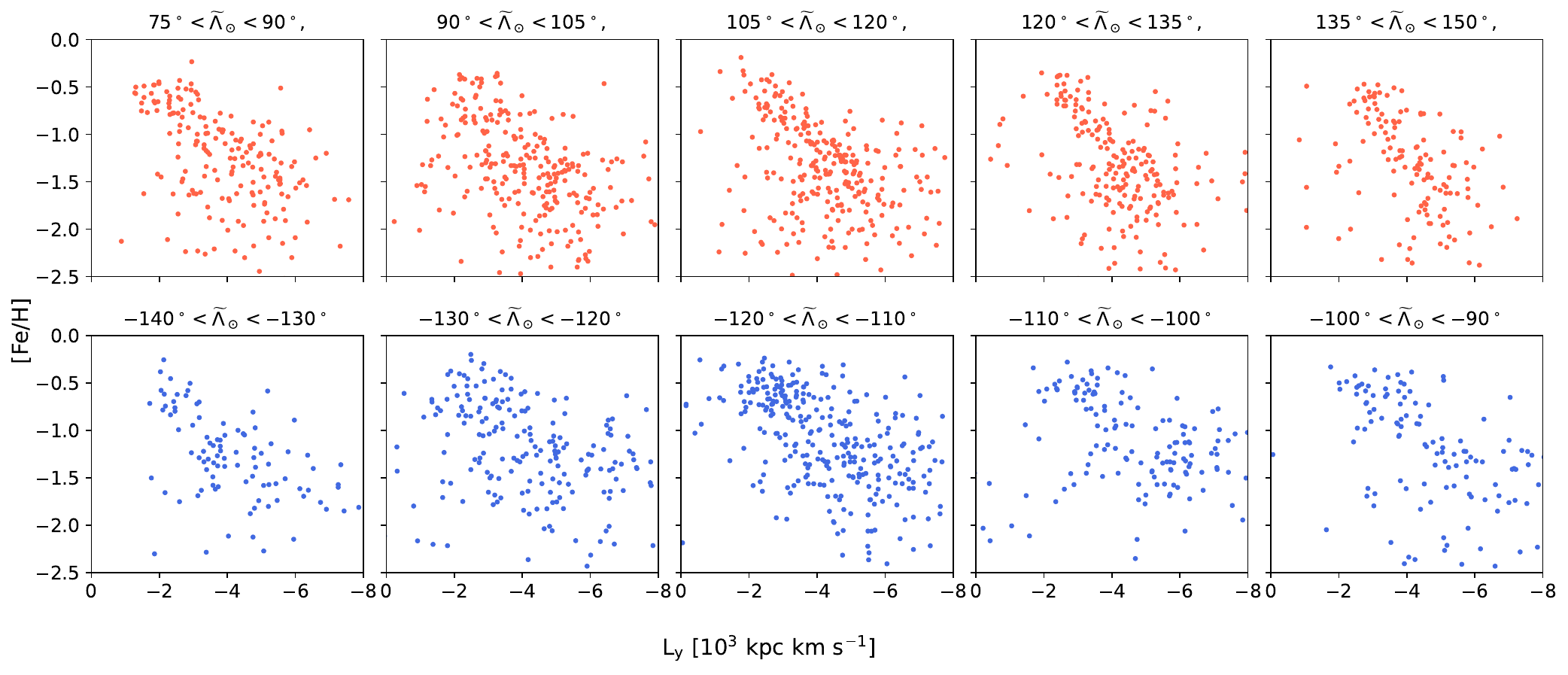}
    \caption{\textbf{ The observed ${\rm[Fe/H]}-L_y$ correlation within individual $\tilde{\Lambda}_\odot$ bins along the Sgr streams.} Top row: Observational data for the leading arm, divided into distinct longitudinal $\tilde{\Lambda}_\odot$ bins. Bottom row: Corresponding observational data for the trailing arm. In both arms, a clear correlation is robustly present within well-sampled  $\tilde{\Lambda}_\odot$ bins.}
    \label{fig_lambda_bin}
\end{figure}

\begin{figure}[h]
    \centering
    \includegraphics[width=13cm,height=6cm]{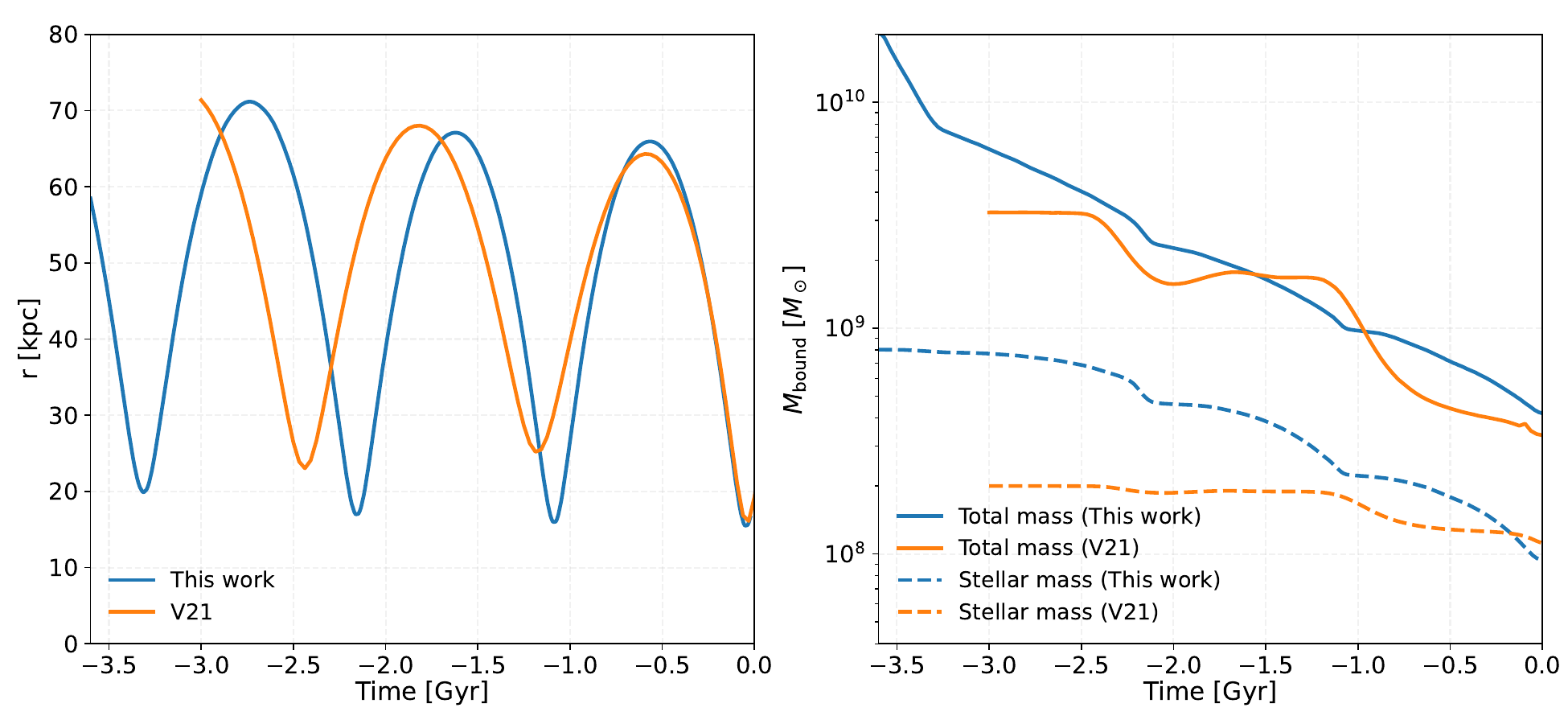}
    \caption{\textbf{Orbital and mass-loss history of the Sgr progenitor.}
    \textbf{Left:} Galactocentric distance of the progenitor as a function of time. Our simulation adopts a longer integration time than V21 \cite{vasiliev_tango_2021}, resulting in an additional early close encounter with the MW.
    \textbf{Right:} Mass evolution of the progenitor over time. Blue and orange lines represent our fiducial model (This work) and the model from V21 \cite{vasiliev_tango_2021}, respectively. Solid lines denote the total mass, while dashed lines track the stellar mass. After the first orbital period, the bound mass in our simulation drops rapidly to $\sim 5\times 10^9\, M_\odot$,  comparable to the initial bound mass adopted in V21. }
    \label{fig_massloss}
\end{figure}

\begin{figure}[h]
    \centering
    \includegraphics[width=14cm,height=13cm]{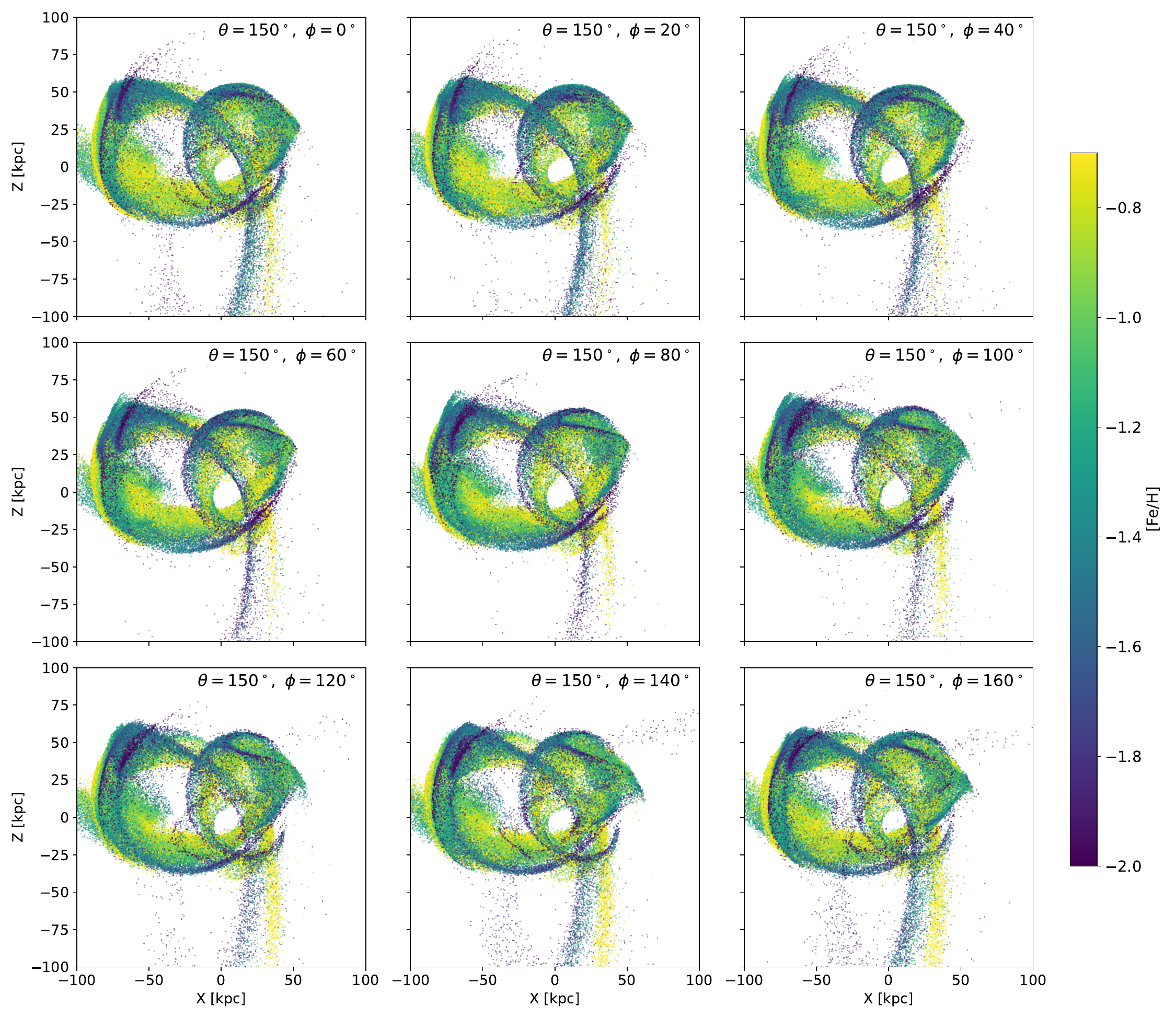}
    \caption{\textbf{Projected spatial distributions of disk particles in the $X-Z$ plane for different orientation angles.} The snapshots vary across different angles of $\phi$ from $0^\circ$ to $160^\circ$ in steps of $20^\circ$, all configured at a fixed $\theta = 150^\circ$. The color encodes the stellar metallicity ${\rm [Fe/H]}$ of each particle. All snapshots are extracted when the simulated Sgr reaches its present-day position. All panels share the same axis limits and color scale to facilitate direct comparison.}\label{supp_fig15}
    \label{fig_xz_diffphi}
\end{figure}

\begin{figure}[h]
    \centering
    \includegraphics[width=8cm,height=8cm]{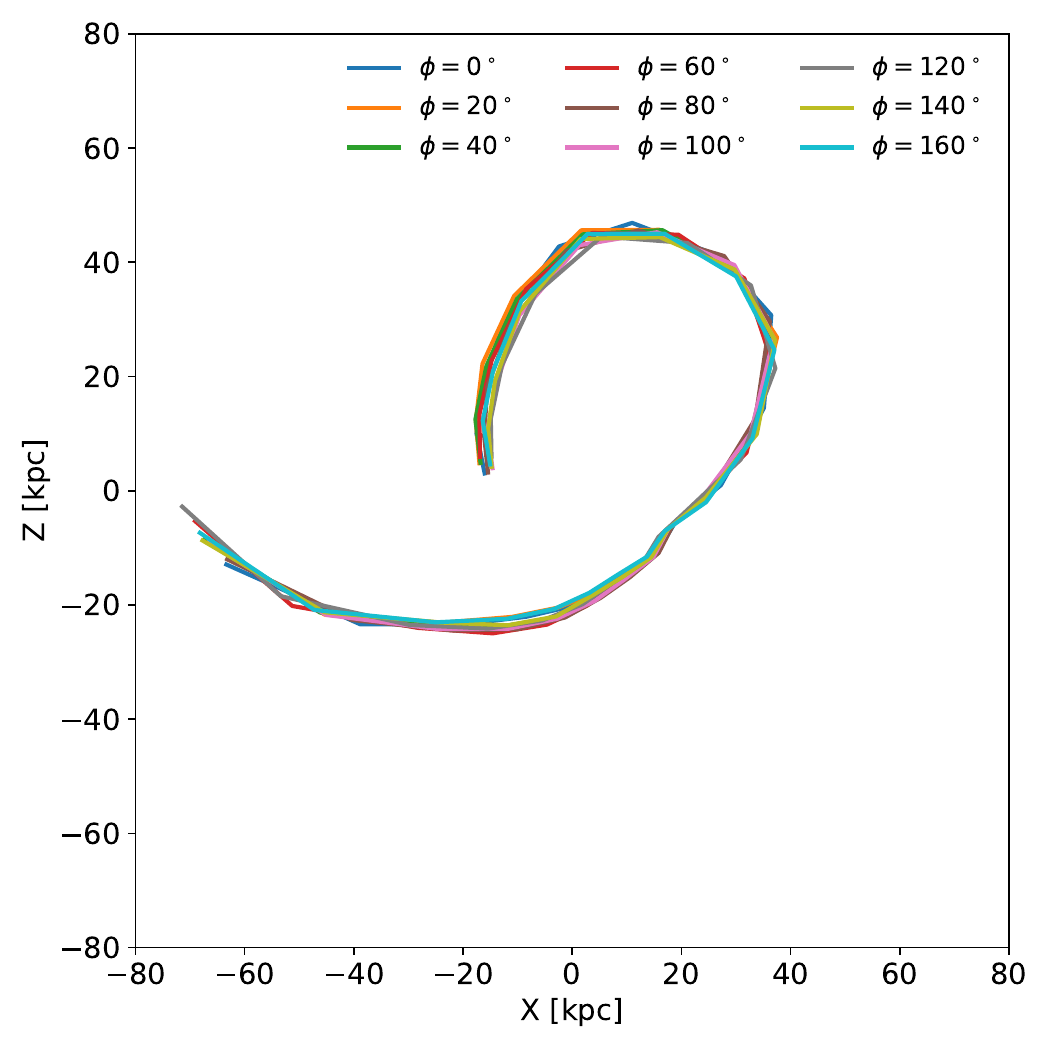}
    \caption{\textbf{Galactocentric $X-Z$ trajectories of the simulated stream for different orientation angles.} The model explores a range of angles for $\phi$ from $0^\circ$ to $160^\circ$ in steps of $20^\circ$, with $\theta$ fixed at $150^\circ$. Each curve corresponds to a different $\phi$ value to illustrate the trajectory variations.}\label{supp_fig16}
    \label{fig_xz_traj_diffphi}
\end{figure}

\begin{figure}[h]
    \centering
    \includegraphics[width=13cm,height=4cm]{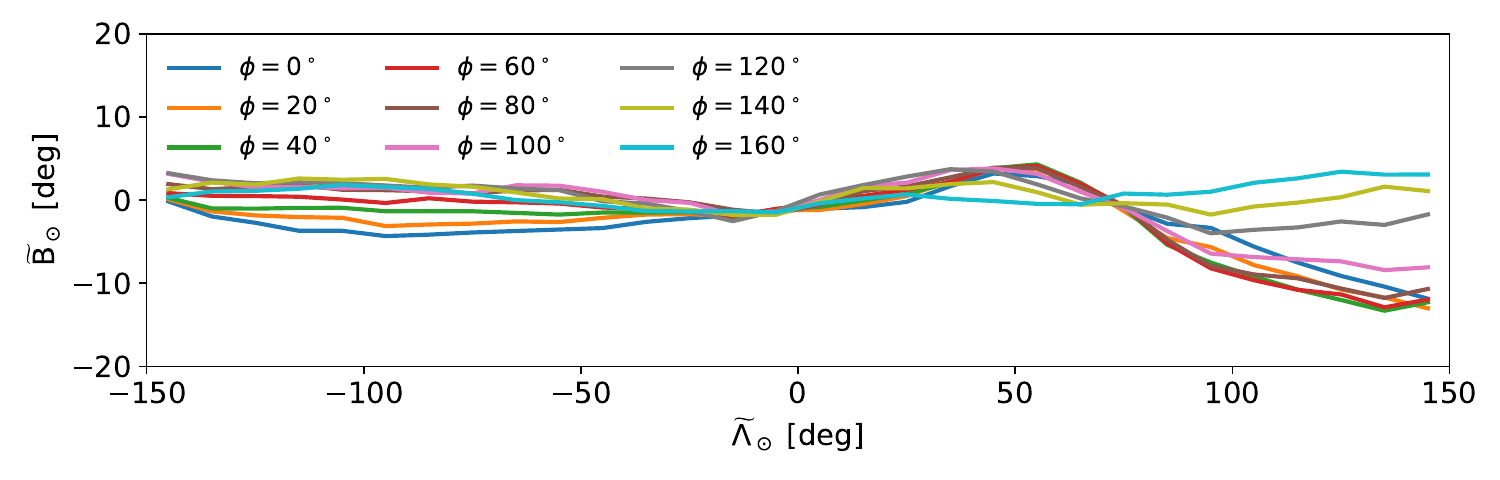}
    \caption{\textbf{Mean stream latitude as a function of longitude for the simulated Sgr streams under different angles.} The parameter space spans different angles of $\phi$ from $0^\circ$ to $160^\circ$ in steps of $20^\circ$, plotted as distinct colored lines, with $\theta$ fixed at $150^\circ$. The relationship highlights the spatial track variations of the stream models.}\label{fig_lam_beta_diffphi}
\end{figure}

\begin{figure}
    \centering
    \includegraphics[width=14cm,height=3.5cm]{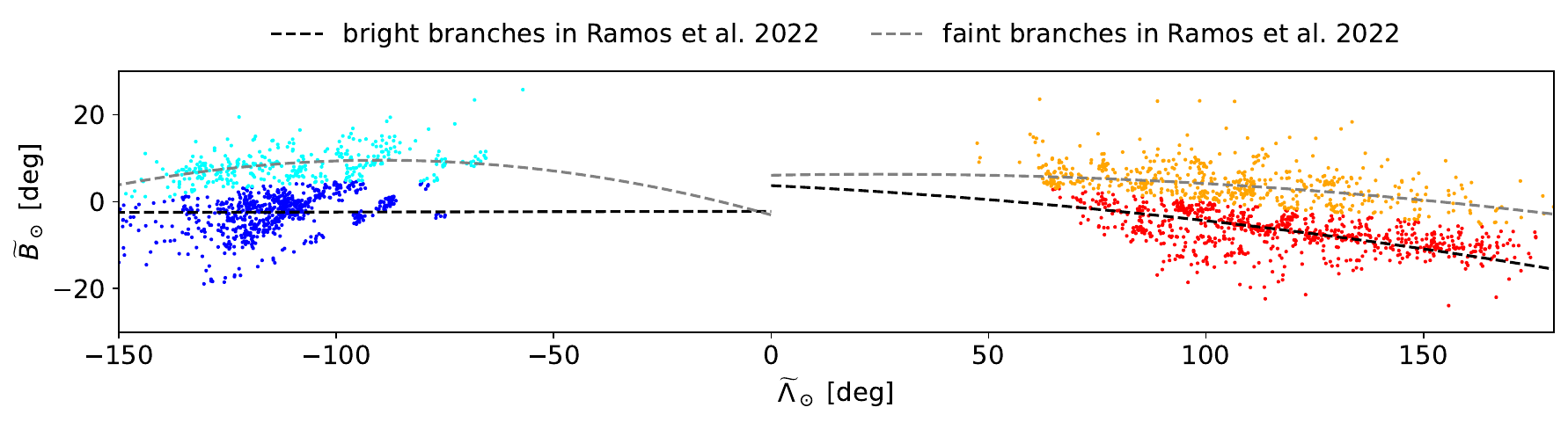}
    \caption{\textbf{Spatial classification of the Sgr stream branches.} The on-sky distribution of Sgr member stars in the stream coordinate frame ($\widetilde{\Lambda}_\odot, \widetilde{B}_\odot$). The sample is decomposed into four components based on the spatial criteria of Ramos et al. (2022)\cite{ramos_2022}: leading-bright (red dots), leading-faint (orange dots), trailing-bright (blue dots), and trailing-faint(cyan dots). For reference, the black and gray dashed lines indicate the approximate location for the bright and faint branches, respectively, as defined by Ramos et al. (2022)\cite{ramos_2022}.}
    \label{lambda_beta}
\end{figure}

\begin{figure}
    \centering
    \includegraphics[width=12cm,height=12cm]{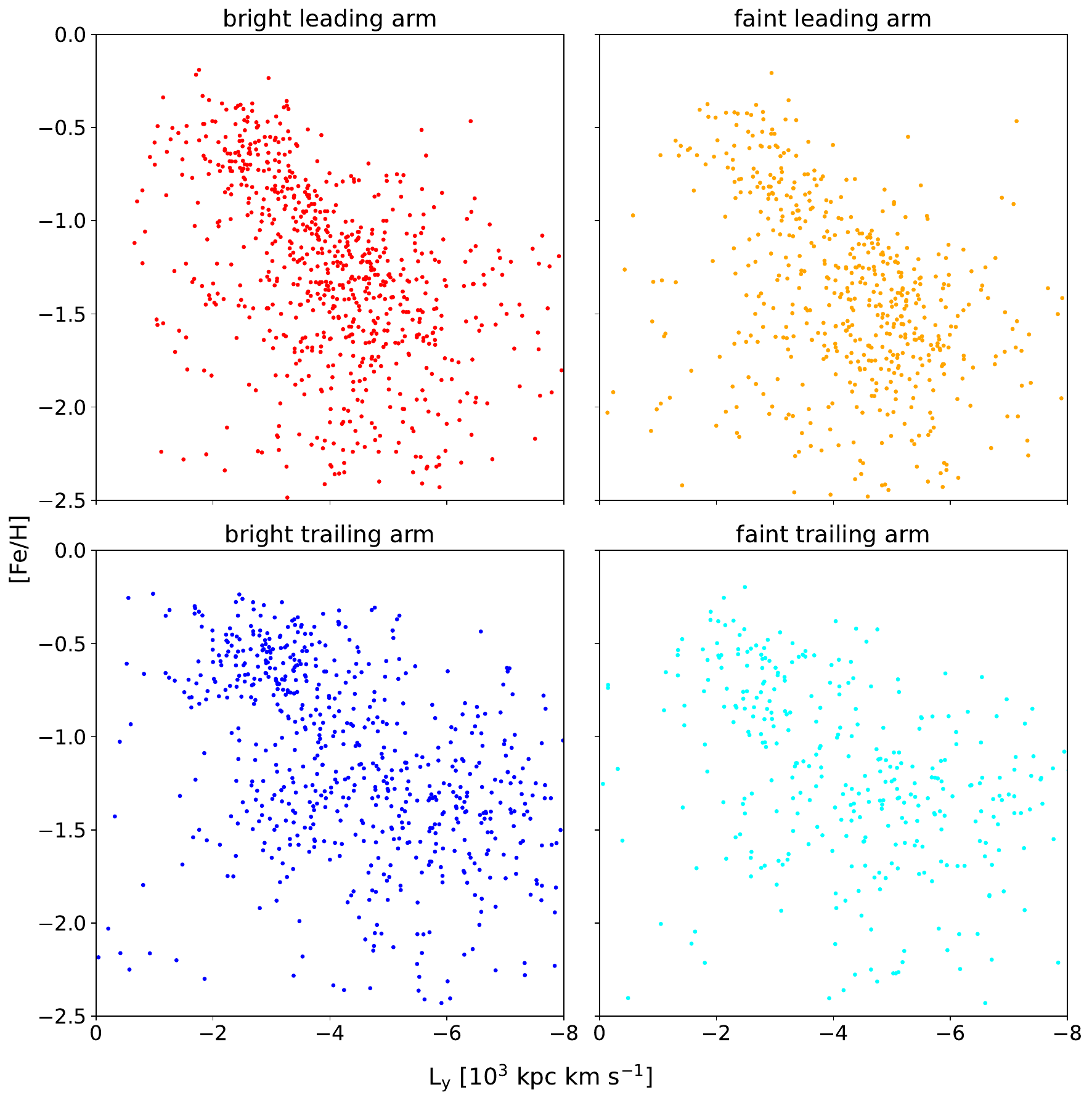}
    \caption{\textbf{Robustness of the ${\rm[Fe/H]}-L_y$ correlation across stream branches.} The ${\rm[Fe/H]}-L_y$ relation is plotted separately for each of the four branches: leading-bright, leading-faint, trailing-bright, and trailing-faint. The characteristic trend is consistently present in each subsample.}
    \label{ly_fe_bifurcation}
\end{figure}

\begin{figure}
    \centering
    \includegraphics[width=13cm,height=7.5cm]{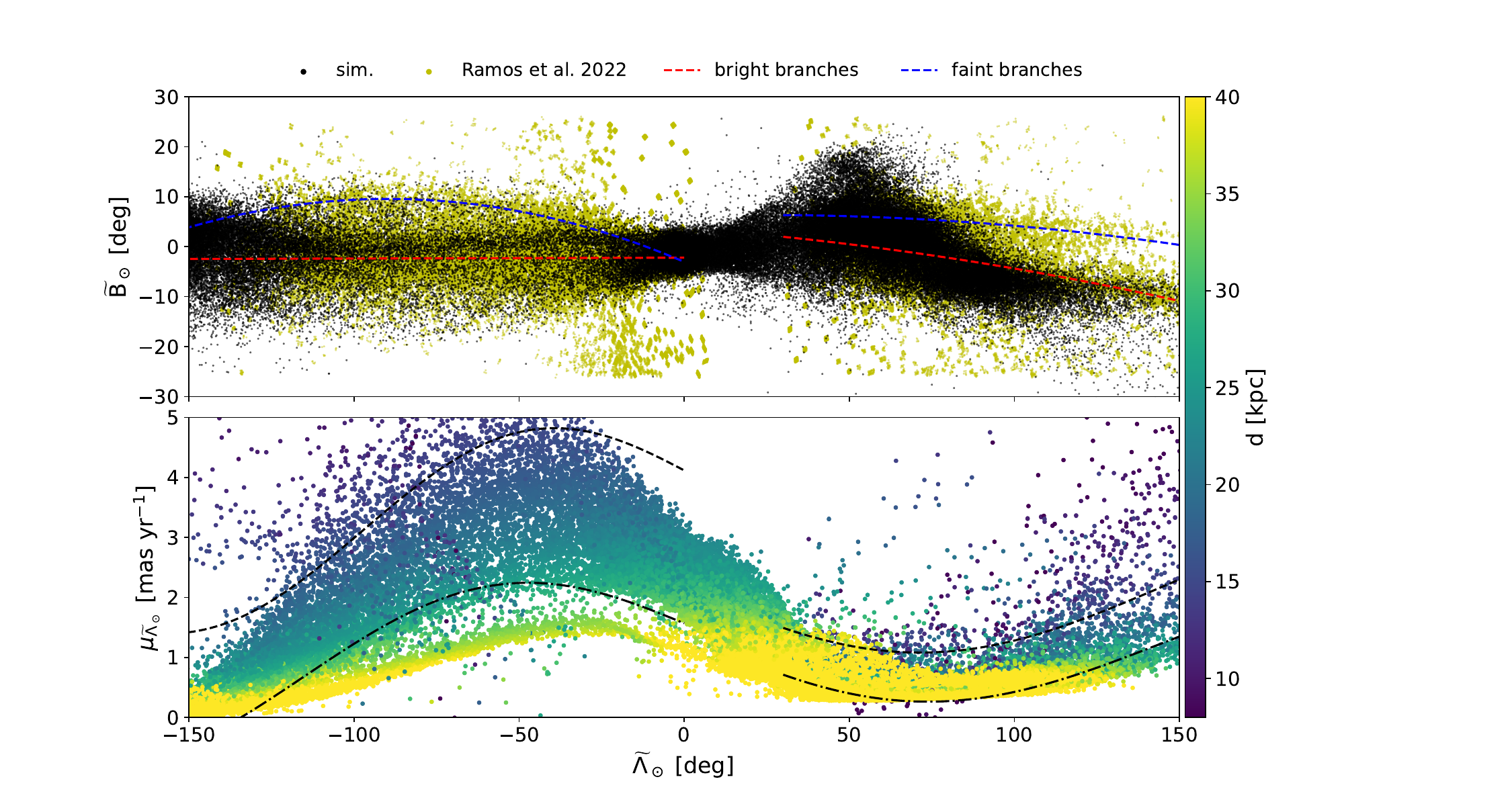}
    \caption{\textbf{Comparison of the fiducial model with observational data}. \textbf{Top:} Spatial distribution in the Sgr coordinate frame. Black points represent stellar particles from the simulation, while yellow points represent observational data from Ramos et al. (2022)\cite{ramos_2022}. The red and blue dashed lines indicate the locations of the bright and faint branches, respectively, as defined by Ramos et al. (2022)\cite{ramos_2022}. \textbf{Bottom:} Tangential proper motion $\mu_{\tilde{\Lambda}\odot}$ as a function of $\tilde{\Lambda}_\odot$ for the simulated stream, colored by heliocentric distance. The lines represent the upper and lower limits of the kinematic selection from Ramos et al. (2022)\cite{ramos_2022}.} 
    \label{fig_compare_ramos}
\end{figure}

\begin{figure}
    \centering
    \includegraphics[width=12cm,height=5cm]{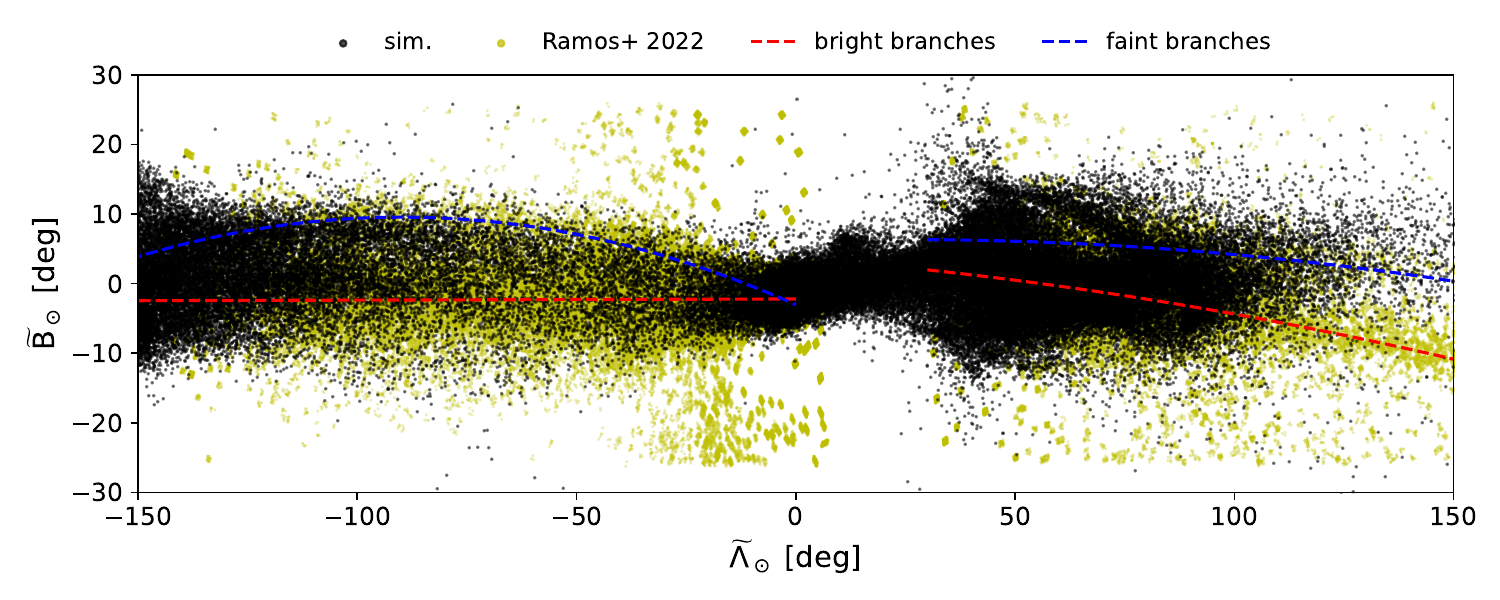}
    \caption{\textbf{Comparison of the simulation ($\theta={150}^\circ,\ \phi={150}^\circ$) and the observational data.} Black points represent stellar particles from the simulation, while yellow points represent observational data from Ramos et al. (2022)\cite{ramos_2022}. The red and blue dashed lines indicate the locations of the bright and faint branches, respectively, as defined by Ramos et al. (2022)\cite{ramos_2022}.}
    \label{fig_sim_bifurcation}
\end{figure}

\begin{figure}
    \centering
    \includegraphics[width=13cm,height=4cm]{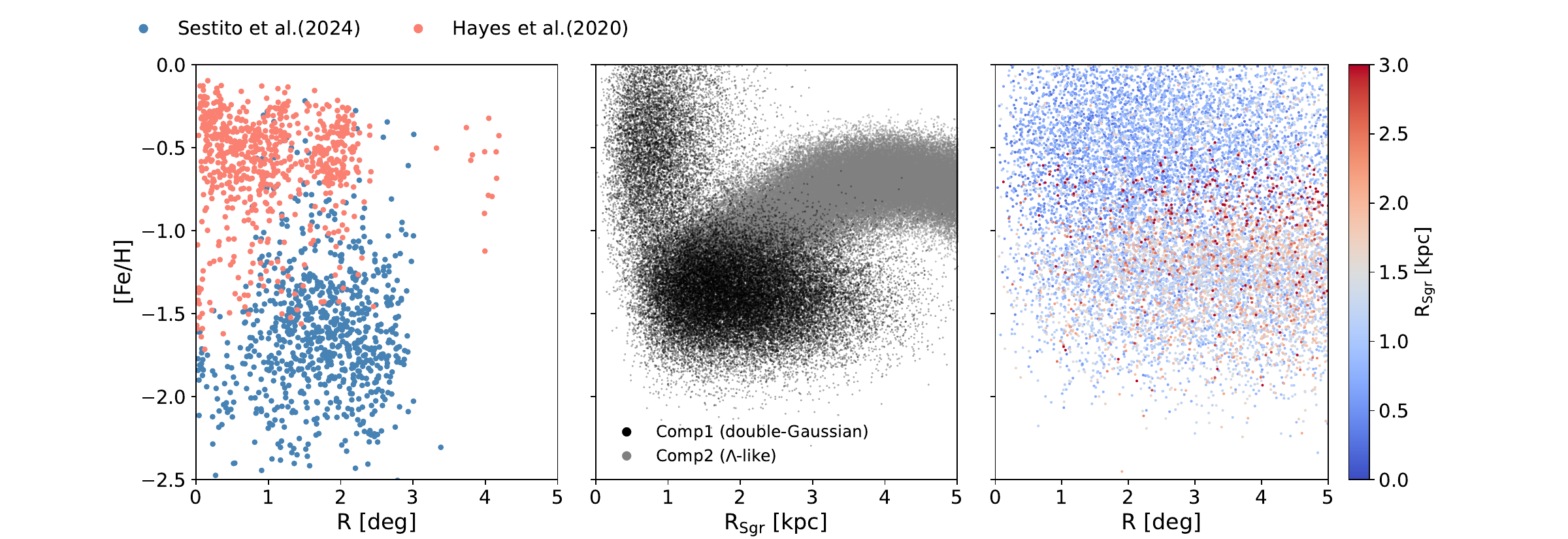}
    \caption{\textbf{Modeling the bimodal metallicity structure of the Sgr core.} \textbf{Left}: Observed metallicity distribution combining data from the Pristine survey (blue; Sestito et al. (2024))\cite{Sestito_2024}) and APOGEE (red; Hayes et al. (2020))\cite{Hayes_2020}, revealing a bimodal metallicity population. \textbf{Middle}: Initial metallicity distribution. $comp.1$ adopts a double-Gaussian profile to explicitly include the metal-rich population, while $comp.2$ follows the $\wedge$-like profile responsible for the ${\rm[Fe/H]}-L_y$ correlation. \textbf{Right}: Simulated present-day metallicity distribution of the core. Points are color-coded by their initial radial position ($\rm{R_{Sgr}}$) in the progenitor.}
    \label{fig_meta_core}
\end{figure}

\begin{figure}
    \centering
    \includegraphics[width=10cm,height=5cm]{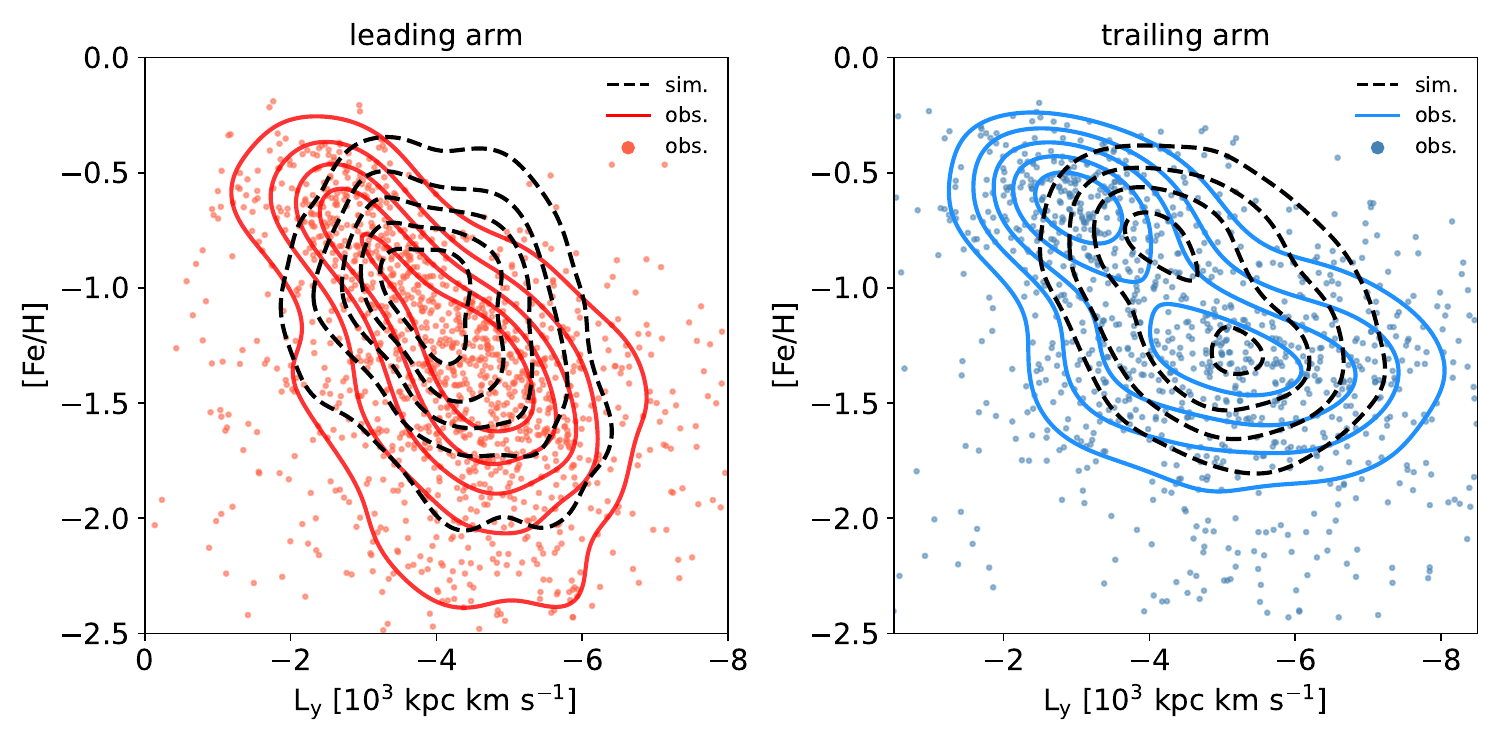}
    \caption{\textbf{The ${\rm[Fe/H]}-L_y$ correlation for a bimodal metallicity Sgr core. }  
    \textbf{Left:} The red contour and red dots represent the observation data and the black contour indicates the ${\rm[Fe/H]}-L_y$ correlation of the leading arm in the simulation. 
    \textbf{Right: } The Blue contour and blue dots represent the observation data, and the black contour indicates the ${\rm[Fe/H]}-L_y$ correlation of the trailing arm in the simulation.}
    \label{fig_core_gradient}
\end{figure}

\begin{figure}
    \centering
    \includegraphics[width=13cm,height=6.5cm]{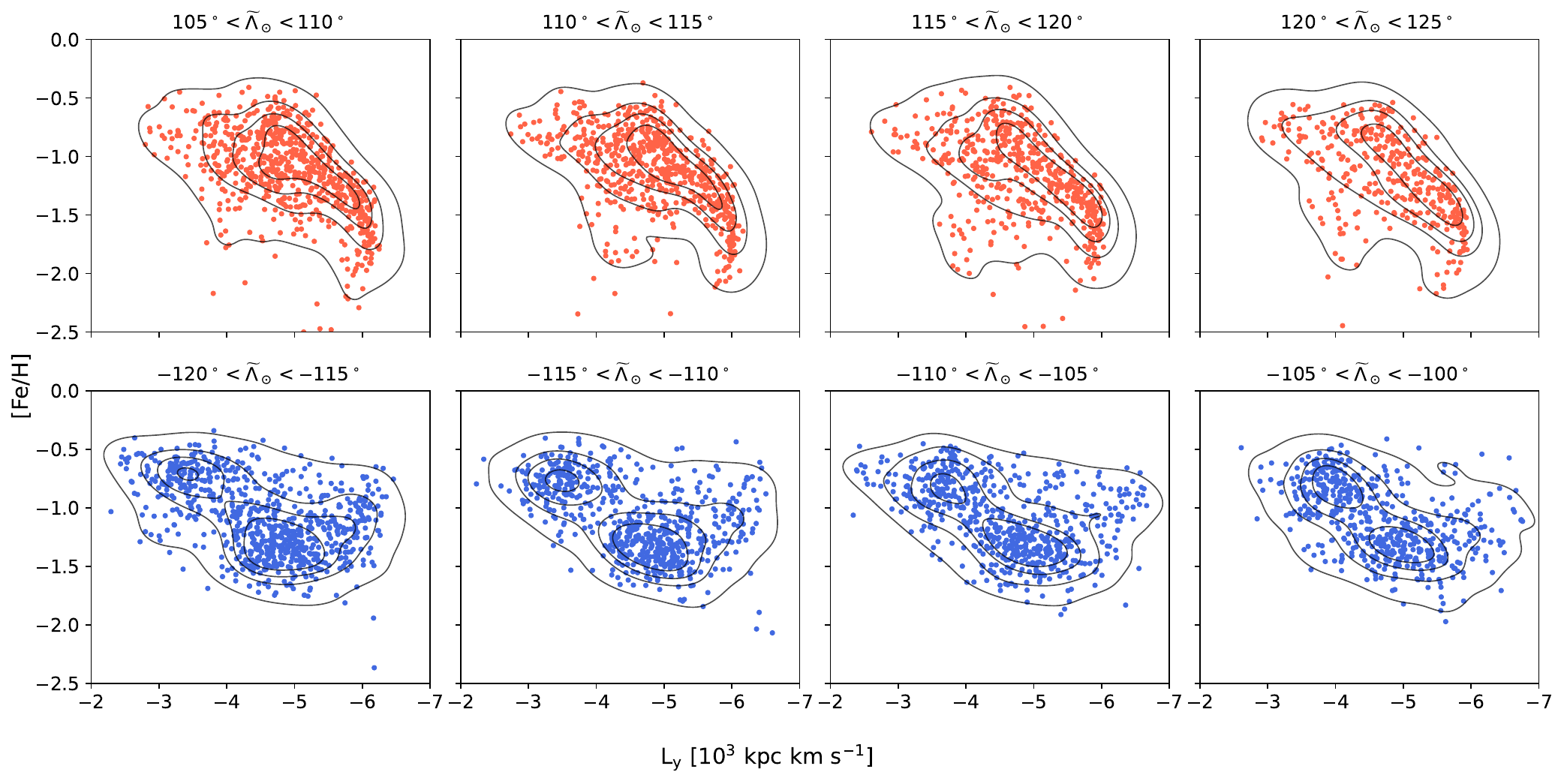}
    \caption{\textbf{The simulated ${\rm[Fe/H]}-Ly$ correlation within $\tilde{\Lambda}_\odot$ bins under the MW+LMC potential.} \textbf{Top row:} Simulated data for the leading arm, focusing specifically on the longitudinal range of $105^\circ < \tilde{\Lambda}_\odot < 125^\circ$, where the observed correlation is significant, divided into distinct $\tilde{\Lambda}\odot$ bins. The individual red dots represent all simulated stellar particles within each bin without the addition of observational errors, while the overlaid black contours outline their corresponding density distributions. 
    \textbf{Bottom row:} Corresponding simulated data for the trailing arm within the specific range of $-120^\circ < \tilde{\Lambda}_\odot < -100^\circ$. In both arms, the ${\rm[Fe/H]}-Ly$ correlation is robustly preserved locally within these individual bins. The individual blue dots show the simulated stellar particles without observational errors, and the black contours depict their underlying density distributions.}
    \label{fig_sim_bin}
\end{figure}

\begin{figure}
    \centering
    \includegraphics[width=13cm,height=8.5cm]{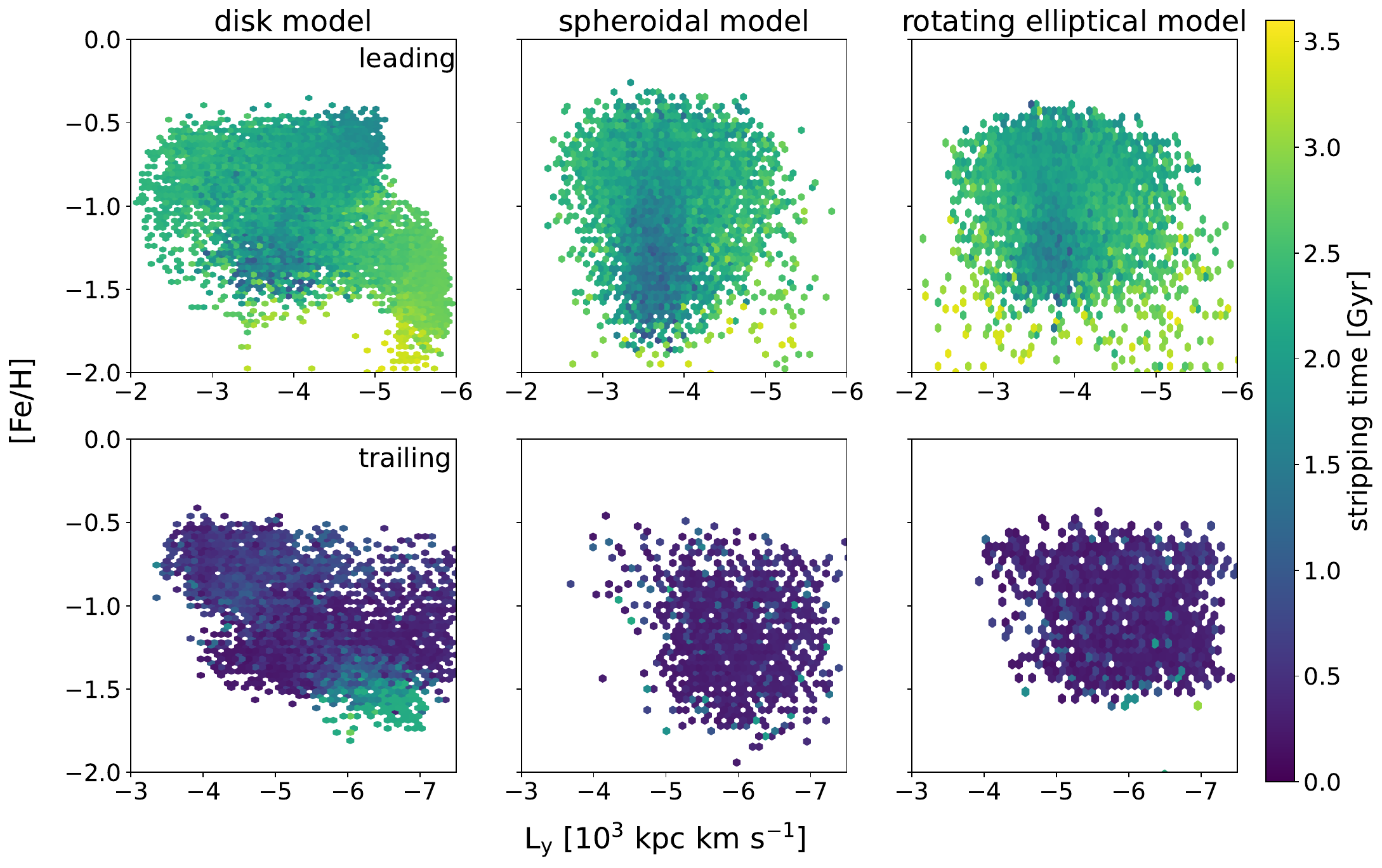}
    \caption{\textbf{Comparison of disk model, spheroidal model and rotating elliptical model. Top row:} Distribution of the leading arm generated by three models in the ${\rm[Fe/H]}-L_y$ space. The disk model adopts our fiducial model. \textbf{Bottom row:} the ${\rm[Fe/H]}-L_y$ correlation of the trailing arms. Each hexagonal bin represents the mean value of all data points within its corresponding region, and color represents the stripped time when the stellar particle left its progenitor. A larger stripped time indicates that the star is stripped earlier.  }
    \label{strip}
\end{figure}

\end{document}